\documentclass[graybox]{svmult}

\usepackage{graphicx}

\usepackage{color}

\usepackage[latin1]{inputenc}
\usepackage{amsfonts}
\usepackage{tikz} 
\usetikzlibrary{matrix}
\usepackage[english]{babel}
\usepackage[latin1]{inputenc}
\usepackage[T1]{fontenc}
\usepackage{ae}

\usepackage{url}

\usepackage{amsmath}

\allowdisplaybreaks[4]
\newcommand\SH{\,\mbox{$\sqcup \! \sqcup$}\,}
\newcommand{\HA}{{\rm H}}
\newcommand{\N}{\mathbb{N}}
\newcommand{\R}{\mathbb{R}}

\newcommand{\abs}[1]{\left|{#1}\right|}

\newcommand{\sign}[1]{\textnormal{sign}\hspace{-0.1em}\left(#1\right)}

\renewcommand{\H}[2]{\textnormal{H}_{#1}\hspace{-0.2em}\left(#2\right)}
\renewcommand{\S}[2]{\textnormal{S}_{#1}\hspace{-0.2em}\left(#2\right)}

\usepackage{color}
\usepackage{tikz}
\usetikzlibrary{matrix}
\usepackage{cite}           
\usepackage{mathptmx}       % selects Times Roman as basic font
\usepackage{helvet}         % selects Helvetica as sans-serif font
\usepackage{courier}        % selects Courier as typewriter font
\usepackage{type1cm}        % activate if the above 3 fonts are
\usepackage{makeidx}         % allows index generation
\usepackage{graphicx}        % standard LaTeX graphics tool
\usepackage{multicol}        % used for the two-column index
\usepackage[bottom]{footmisc}% places footnotes at page bottom
\newcommand{\Li}{{\rm Li}}
\newcommand{\Mvec}{{\rm \bf M}}
\newcommand{\ds}{\displaystyle}
\usepackage{amsmath,amssymb}

\makeindex             % used for the subject index
\begin{document}

\title*{{\footnotesize{\sf DESY 13-073, DO-TH 13/10, SFB/CPP-13-26, LPN13-025}}\\
Harmonic Sums\index{Harmonic Sums}, Polylogarithms\index{Polylogarithms}, 
Special Numbers\index{Special Numbers}, and their Generalizations}
\author{Jakob Ablinger and Johannes Bl\"umlein}
\institute{Jakob Ablinger,\\  Research Institute for Symbolic Computation (RISC)
Johannes Kepler University, Altenbergerstra\ss{}e 69, A-4040 Linz, Austria,
\email{jablinge@risc.uni-linz.ac.at},\\
Johannes Bl\"umlein, \\ Deutsches Elektronen-Synchrotron, DESY, Platanenallee 6, D-14738
Zeuthen, Germany,
\email{Johannes.Bluemlein@desy.de}}
\maketitle

%%%%%%%%%%%%%%%%%%%%%%%%%%%%%%%%%%%%%%%%%%%%%%%%%%%%%%%%%%%%%%%%%%%%%%%%%%%%%%%%%%%%%%%%%%%%%%%
\abstract{
In these introductory lectures we discuss classes of presently known nested sums, associated 
iterated integrals\index{Iterated integrals}, and special constants which hierarchically appear in 
the evaluation of massless and massive Feynman diagrams\index{Feynman diagrams} at higher loops. 
These quantities are elements of stuffle\index{stuffle algebra} and shuffle 
algebras\index{shuffle algebra} implying algebraic relations being widely independent of the 
special quantities considered. They are supplemented by structural 
relations\index{
structural relations}. The generalizations are given in terms of generalized 
harmonic sums, (generalized) cyclotomic sums, and sums containing in addition binomial and 
inverse-binomial weights. To all these quantities iterated integrals and special numbers 
are associated. We also discuss the analytic continuation\index{analytic continuation} of nested 
sums\index{nested sum} of different kind to complex values of the external summation bound~$N$.}
%%%%%%%%%%%%%%%%%%%%%%%%%%%%%%%%%%%%%%%%%%%%%%%%%%%%%%%%%%%%%%%%%%%%%%%%%%%%%%%%%%%%%%%%%%%%%%%

%%%%%%%%%%%%%%%%%%%%%%%%%%%%%%%%%%%%%%%%%%%%%%%%%%%%%%%%%%%%%%%%%%%%%%%%%%%%%%%%%%%%%%%%%%%%%%%
\section{Introduction}
\label{sec:1}
%%%%%%%%%%%%%%%%%%%%%%%%%%%%%%%%%%%%%%%%%%%%%%%%%%%%%%%%%%%%%%%%%%%%%%%%%%%%%%%%%%%%%%%%%%%%%%%

\vspace{1mm}\noindent
In the solution of physical problems very often new classes of special 
functions\index{special functions} have been created during the last three centuries, 
cf.~\cite{LENSE,SOMMERFELD,KAMKE,KRATZER,ANDREWS}. This applies especially also to the analytic 
calculation of Feynman-parameter integrals \cite{NAKANISHI} for massless and massive two- and 
more-point functions, also containing local operator\index{local operator} insertions and 
corresponding quantities, cf.~\cite{Vermaseren:2005qc,Bierenbaum:2009mv}. In case of zero 
mass-scale quantities the associated integrals map to special numbers, lately having been called 
periods\index{periods} \cite{KONTSEVICH}, see also \cite{Bogner:2007mn}. In case of single-scale 
quantities, expressed as a ratio $x \in [0,1]$ to the defining mass scale, the integrals are 
Poincar\'e iterated integrals\index{Poincar\'e iterated integral} \cite{POINC,POLYLOG2} 
or they emerge as a
Mellin-transform\index{Mellin transform} at $N \in \mathbb{N}$ \cite{MELLIN} in terms of 
multiply nested sums. A systematic way to these structures has been described in
\cite{Blumlein:2009ta,Blumlein:2010zv}. Here an essential tool consists in representations 
by Mellin--Barnes \cite{MELLINBARNES} integrals. They are applicable also for integrals of 
multi-scale more-loop and multi-leg Feynman integrals \cite{Gluza:2007rt}, which are, however, 
less explored at present. 

In the practical calculations dimensional regularization\index{dimensional regularization} in 
$D = 4 + \varepsilon$ space-time dimensions \cite{DIMREG} is used, which is essential to maintain 
conservation laws due to the Noether theorem and probability. It provides the 
singularities\index{singularity} of the problem  in terms of poles in $\varepsilon$. However, the 
Feynman parameter integrals are not performed over rational integrands but 
hyperexponential\index{hyperexponential} ones. Thus  one passes through higher transcendental 
functions \cite{KRATZER,ANDREWS} from the beginning. The renormalization is carried out in the 
$\overline{\rm MS}$-scheme, chosen as the standard. In new calculations various ingredients as 
anomalous dimensions\index{anomalous dimension} and expansion coefficients of the 
$\beta$-functions\index{beta function} needed in the renormalization\index{renormalization} 
can thus be used referring to results given in the literature. At higher orders the calculation 
of these quantities requests a major investment and is not easily repeated at present within other 
schemes in a short time.

With growing complexity of the perturbative calculations in Quantum Field Theories the 
functions emerging in integration and summation had to be systematized. While a series of massless
2--loop calculations, cf.~\cite{TL}, during the 1980ies and 90ies initially still could be performed 
referring to the classical polylogarithms \cite{DILOG,SPENCE,POLYLOG1,POLYLOG2,POLYLOG3} and 
Nielsen-integrals \cite{NIELSEN}, the structure of the results became readily involved. In 1998 a 
first general standard was introduced \cite{Vermaseren:1998uu,Blumlein:1998if} by the nested 
harmonic sums, and shortly after the harmonic polylogarithms \cite{Remiddi:1999ew}. Further 
extensions are given by the generalized harmonic sums, the so-called 
S-sums \cite{Moch:2001zr,Ablinger:2013cf} and the (generalized) cyclotomic sums  
\cite{Ablinger:2011te}, see~Figure~1. Considering problems at even higher loops and a growing 
number of legs, also associated with more mass scales, one expects various new levels of 
generalization to emerge. In particular, also elliptic integrals will contribute \cite{ELLIPTIC}.
%-----------------------------------------------------------------------------------------------------
\def\firstcircle{(0,0) circle (3.5cm and 1.5cm)}
\def\secondcircle{(0:4cm) circle (3.5cm and 1.5cm)}
\def\thirdcircle{(0:2cm) circle (6cm and 2.5cm)}

\colorlet{circle area}{green!20}

\tikzset{filled/.style={fill=circle area, draw=circle edge, thick},
    outline/.style={draw=circle edge, thick}}
\colorlet{circle edge}{black!100}
\colorlet{circle area}{black!20}
\setlength{\parskip}{5mm}
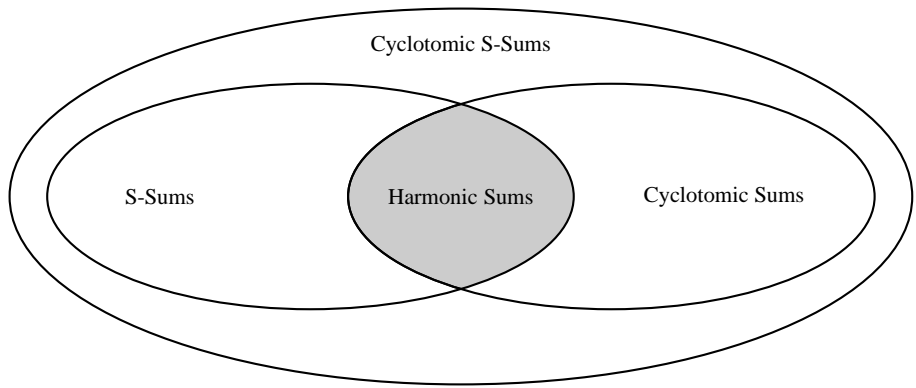
\begin{figure}
\centering
\begin{tikzpicture}
     \begin{scope}
         \clip \firstcircle;
         \fill[filled] \secondcircle;
     \end{scope}
     \draw[outline] \firstcircle;
     \draw[outline] \secondcircle;
     \draw[outline] \thirdcircle;
     \draw (2,0) node {Harmonic Sums};
     \draw (-2,0) node {S-Sums};
     \draw (5.5,0) node {Cyclotomic Sums};
     \draw (2,2) node {Cyclotomic S-Sums};
\end{tikzpicture}
\begin{center}
\caption{\label{connectionfigure}Relations between the different extensions of harmonic sums.}
\end{center}
\end{figure}
%-----------------------------------------------------------------------------------------------------
These structures can be found systematically by applying symbolic summation, cf.~\cite{SIGMA}, and
integration formalisms, cf.~\cite{BRONSTEIN,RAAB}, which also allow to proof the relative 
transcendence of the basis elements found and are therefore applied in the calculation
of Feynman diagrams.

In this survey we present an introduction to a series of well-studied structures which have been 
unraveled during the last years.
The paper is organized as follows. In Section~2 a survey is given on polylogarithms, Nielsen 
integrals and harmonic polylogarithms. In Section~3 harmonic sums are discussed. Both harmonic
polylogarithms and harmonic sums obey algebraic and structural relations on which a survey is 
given in Section~4. In Section~5 we discuss properties of the multiple zeta values which emerge
as special constants in the context of harmonic sums and polylogarithms. 
The S-sums, associated iterated integrals, and special numbers are considered in 
Section~6. The generalization of harmonic sums and S-sums to (generalized) cyclotomic sums,
polylogarithms and numbers is given in Section~7. A further generalization, which appears in 
massive multi-loop calculations, to nested binomial and inverse-binomial harmonic sums and 
polylogarithms is outlined in Section~8. Finally, we discuss in Section~9 the analytic 
continuation of the different kind of nested sums in the argument $N$ to complex numbers, 
which is needed in various physical applications. Section~10 contains the conclusions.
The various mathematical relations between the different quantities being discussed in the 
present article are implemented in the package {\tt HarmonicSums.m} \cite{Ablinger:2013cf,JA1}.
%%%%%%%%%%%%%%%%%%%%%%%%%%%%%%%%%%%%%%%%%%%%%%%%%%%%%%%%%%%%%%%%%%%%%%%%%%%%%%%%%%%%%%%%%%%%%%%
\section{Polylogarithms, Nielsen Integrals, Harmonic Polylogarithms}
\label{sec:2}
%%%%%%%%%%%%%%%%%%%%%%%%%%%%%%%%%%%%%%%%%%%%%%%%%%%%%%%%%%%%%%%%%%%%%%%%%%%%%%%%%%%%%%%%%%%%%%%

\vspace{1mm}\noindent
Different particle propagators\index{propagator} $1/A_k(p_i,m_i)$ can be linked using Feynman's 
integral 
representation \cite{Feynman:1949zx}
%---------------------------------------------------------------------------------------------
\begin{eqnarray}
\frac{1}{A_1^{\nu_1} ... A_n^{\nu_n}} = \frac{\Gamma(\sum_{k=1}^n 
\nu_k)}{\prod_{k=1}^n \Gamma(\nu_k)} 
\int_0^1 \prod_{k=1}^n dx_k \frac{\prod_{k=1}^n x^{\nu_k-1}}
{\left(\sum_{k=1}^n x_k A_k\right)^{\sum_{k=1}^n \nu_k}  
}\delta\left(1 - \sum_{k=1}^n x_k\right)~, \nu_i \in \mathbb{R}. 
\end{eqnarray}
%---------------------------------------------------------------------------------------------
While the momentum integrals over $p_i$ can be easily performed, the problem consists in
integrating the Feynman parameters $x_k$. In the simplest cases the associated integrand is
a multi-rational function. In the first integrals one obtains multi-rational functions, but also
logarithms \cite{BRONSTEIN}. The logarithms\index{logarithm} \cite{NAPIER} have to be introduced 
as new functions
being transcendental to the rational functions
%---------------------------------------------------------------------------------------------
\begin{eqnarray}
\int_0^x \frac{dz}{1-z} = - \ln(1-x),~~\text{etc.}
\end{eqnarray}
%---------------------------------------------------------------------------------------------
Iterating this integral by
%---------------------------------------------------------------------------------------------
\begin{eqnarray}
\int_0^x \frac{dz_1}{z_1} \int_0^{z_1} \frac{dz_2}{1-z_2} = \Li_2(x)
\end{eqnarray}
%---------------------------------------------------------------------------------------------
one obtains the dilogarithm\index{dilogarithm} or Spence-function \cite{DILOG,SPENCE}, which may 
be extended to the
classical polylogarithms \cite{SPENCE,POLYLOG1,POLYLOG2,POLYLOG3}
%---------------------------------------------------------------------------------------------
\begin{eqnarray}
\int_0^x \frac{dz}{z} \Li_{n-1}(z) = \Li_n(x), n \in \mathbb{N}~.
\end{eqnarray}
%---------------------------------------------------------------------------------------------
All these functions are transcendental to the former ones. For an early occurrence of the dilogarithm
in Quantum Field Theory see \cite{RACAH}.

The above iterations are special 
cases in iterating differential forms in $\{dz/z, dz/(1-z)\}$. The general case is described 
as Nielsen integrals\index{Nielsen integral} \cite{NIELSEN}.
%---------------------------------------------------------------------------------------------
\begin{eqnarray}
\label{eq:NIELS}
{\sf S}_{n,p}(x) = \frac{(-1)^{n+p-1}}{(n-1)! p!} \int_0^1 \frac{dz}{z} \ln^{n-1}(z) \ln^p(1-xz)~.
\end{eqnarray}
%---------------------------------------------------------------------------------------------
Likewise, one might also consider the set $\{dz/z, dz/(1+z)\}$. Nielsen integrals obey the 
relation
%---------------------------------------------------------------------------------------------
\begin{eqnarray}
{\sf S}_{n-1,p}(x) = \frac{d}{dx} {\sf S}_{n,p}(x)~. 
\end{eqnarray}
%---------------------------------------------------------------------------------------------
One may derive serial representations around $x=0$, as e.g.:
%---------------------------------------------------------------------------------------------
\begin{eqnarray}
\Li_n(x)   = \sum_{k=1}^\infty \frac{x^k}{k^n},~~
{\sf S}_{1,2}(x) = \sum_{k=2}^\infty \frac{x^k}{k^2} S_1(k-1),~~
{\sf S}_{2,2}(x) = \sum_{k=2}^\infty \frac{x^k}{k^3} S_1(k-1)~,
\end{eqnarray}
%---------------------------------------------------------------------------------------------
see also~\cite{Fleischer:1998nb}.
Here $S_1(n) = \sum_{k=1}^n (1/k)$ denotes the harmonic sum. The Nielsen integrals obey various 
relations \cite{DILOG,SPENCE,POLYLOG1,POLYLOG2,POLYLOG3,NIELSEN}. A few examples are~:
%---------------------------------------------------------------------------------------------
\begin{eqnarray}
\label{eq:SP1}
\Li_2(1-x) &=& - \Li_2(x) - \ln(x) \ln(1-x) + \zeta_2\\
\label{eq:a1}
\Li_2\left(-\frac{1}{x}\right) &=& - \Li_2(-x) - \frac{1}{2} \ln^2(x) - \zeta_2\\
\label{eq:a2}
\Li_3(1-x) &=& -{\sf S}_{1,2}(x) - \ln(1-x) \Li_2(x) - \frac{1}{2} \ln(x) \ln^2(1-x) + \zeta_2 \ln(1-x)
+ \zeta_3 \nonumber\\
\\
\Li_4\left( - \frac{x}{1-x}\right) &=& \ln(1-x)[\Li_3(x)-{\sf S}_{1,2}(x)] + {\sf S}_{2,2}(x) - \Li_4(x) 
- {\sf S}_{1,3}(x) 
\nonumber\\ &&
- \frac{1}{2} \ln^2(1-x) \Li_2(x) - \frac{1}{24} \ln^4(1-x) \\
\Li_n(x^2) &=& 2^{n-1} \left[\Li_n(x) + \Li_n(-x)\right]\\
\Li_2(z) &=& \frac{1}{n} \sum_{x^n = z} \Li_2(x), n \in \mathbb{N} \backslash \{0\} \\
\label{eq:SP2}
{\sf S}_{2,2}(1-x) &=& -{\sf S}_{2,2}(x) + \ln(x) {\sf S}_{1,2}(x) 
- \left[\Li_3(x) - \ln(x) \Li_2(x) - \zeta_3 \right] \ln(1-x) 
\nonumber\\ &&
+ \frac{1}{4} \ln^2(x) \ln^2(1-x)
+ \frac{\zeta_4}{4}~. 
\end{eqnarray}
%---------------------------------------------------------------------------------------------
Here $\zeta_n = \sum_{k=1}^\infty (1/k^n),~~n \geq 2, n \in \mathbb{N}$ are values
of Riemann's $\zeta$--function\index{Zeta function}.

Going to higher orders in perturbation theory it turns out that the Nielsen integrals are 
sufficient for massless and some massive two-loop problems, cf.~\cite{Blumlein:1998if,TWOLOOP}, 
and as well for the 3-loop anomalous dimensions \cite{ANDI3}, allowing for some extended arguments 
as $-x, x^2$. At a given level of complexity, however, one has to refer to a more general 
alphabet, namely 
%---------------------------------------------------------------------------------------------
\begin{eqnarray}
\label{eq:ALPH1}
\mathfrak{A} = \{\omega_0, \omega_1, \omega_{-1}\} \equiv \{dz/z, dz/(1-z), 
dz/(1+z)\}~. 
\end{eqnarray}
%---------------------------------------------------------------------------------------------
The corresponding iterated integrals are called harmonic polylogarithms (HPLs) 
\cite{Remiddi:1999ew}. Possibly the first new integral is
%---------------------------------------------------------------------------------------------
\begin{eqnarray}
\HA_{-1,0,0,1}(x) = \int_0^x \frac{dz}{z} \frac{\Li_3(z)}{1+z}~.
\end{eqnarray}
%---------------------------------------------------------------------------------------------
Here we use a systematic notion defining the Poincar\'e iterated integrals \cite{POINC,POLYLOG2}, 
unlike the 
case in (\ref{eq:NIELS}). The 
weight $w = 1$ HPLs are
%---------------------------------------------------------------------------------------------
\begin{eqnarray}
\HA_{0}(x) = \ln(x),~~~
\HA_{1}(x) = -\ln(1-x),~~~
\HA_{-1}(x) = \ln(1+x),
\end{eqnarray}
%---------------------------------------------------------------------------------------------
with the definition of $\HA_{0,...,0}(x) = \ln^n(x)/n!$ for all $x$ indices equal to zero.
The above functions have the following representation
%---------------------------------------------------------------------------------------------
\begin{eqnarray}
\Li_n(x) = \int_0^x \omega_0^{n-1} \omega_1,~~ 
{\sf S}_{p,n}(x) = \int_0^x \omega_0^{p} \omega_1^n,~~ 
\HA_{\vec{m}_w}(x) = \int_0^x \prod_{l=1}^k \omega_{m_l}~,
\end{eqnarray}
%---------------------------------------------------------------------------------------------
where the corresponding products are non-commutative, $\vec{m}_w$ is of length $k$ and
$x \geq z_1 \geq ... \geq z_m$.

Harmonic polylogarithms obey algebraic and structural relations, which will be discussed
in Section~\ref{sec:4}. Numerical representations of HPLs were given in 
\cite{Gehrmann:2001pz,Vollinga:2004sn}.
%%%%%%%%%%%%%%%%%%%%%%%%%%%%%%%%%%%%%%%%%%%%%%%%%%%%%%%%%%%%%%%%%%%%%%%%%%%%%%%%%%%%%%%%%%%%%%%
\section{Harmonic Sums}
\label{sec:3}
%%%%%%%%%%%%%%%%%%%%%%%%%%%%%%%%%%%%%%%%%%%%%%%%%%%%%%%%%%%%%%%%%%%%%%%%%%%%%%%%%%%%%%%%%%%%%%%

\vspace{1mm}\noindent
The harmonic sums are recursively defined by
%----------------------------------------------------------------------------------------------
\begin{eqnarray}
S_{b,\vec{a}}(N) &=& \sum_{k=1}^N \frac{({\sf sign}(b))^k}{k^{|b|}}
S_{\vec{a}}(k)~,~~~~S_{\emptyset}(N) = 1~,~~~b, a_i \in {\mathbb Z} \backslash \{0\}.
\end{eqnarray}
%----------------------------------------------------------------------------------------------
In physics applications they appeared early in \cite{GonzalezArroyo:1979df,Mertig:1995ny}.
Their systematic use dates back to 
Refs.~\cite{Vermaseren:1998uu,Blumlein:1998if}. They can be
represented as a Mellin transform
%---------------------------------------------------------------------------------------------
\begin{eqnarray}
\label{eq:MEL}
S_{\vec{a}}(N) =
\Mvec\left[f(x)\right](N) = \int_0^1 dx~x^{N-1}~f(x),~~N \in \mathbb{N} \backslash \{0\},
\end{eqnarray}
%---------------------------------------------------------------------------------------------
where $f(x)$ denotes a linear combination of HPLs. 
For example, 
%---------------------------------------------------------------------------------------------
\begin{eqnarray}
S_{-2,1,1}(N) &=& (-1)^{N+1} \int_0^1 dx \frac{\HA_{0,1,1}(x) - \zeta_3}{x+1}
-\Li_4\left(\frac{1}{2}\right)
-\frac{\ln^4(2)}{24}
+\frac{\ln^2(2) \zeta_2}{4}
\nonumber\\ &&
-\frac{7 \ln(2) \zeta_3}{8}
+\frac{\zeta_2^2}{8}
\end{eqnarray}
%---------------------------------------------------------------------------------------------
holds.
Harmonic sums possess algebraic and structural
relations, cf.~Sect.~\ref{sec:4}. 
In the limit $N \rightarrow \infty$ they define the multiple 
zeta values, cf.~Sect.~\ref{sec:5}. They are originally defined at integer argument $N$.
In physical applications they emerge in the context of the light--cone expansion \cite{LCE}.
The corresponding operator matrix elements are analytically continued to complex values of $N$
either from the even {\it or} the odd integers, cf. Sect.~\ref{sec:9}.
%%%%%%%%%%%%%%%%%%%%%%%%%%%%%%%%%%%%%%%%%%%%%%%%%%%%%%%%%%%%%%%%%%%%%%%%%%%%%%%%%%%%%%%%%%%%%%%
\section{Algebraic and Structural Relations}
\label{sec:4}
%%%%%%%%%%%%%%%%%%%%%%%%%%%%%%%%%%%%%%%%%%%%%%%%%%%%%%%%%%%%%%%%%%%%%%%%%%%%%%%%%%%%%%%%%%%%%%%
\subsection{Algebraic Relations}
%%%%%%%%%%%%%%%%%%%%%%%%%%%%%%%%%%%%%%%%%%%%%%%%%%%%%%%%%%%%%%%%%%%%%%%%%%%%%%%%%%%%%%%%%%%%%%%

\vspace{1mm}\noindent
Algebraic relations of harmonic polylogarithms and harmonic sums, respectively, are implied by
their products and depend on their index structure only, i.e. they are a consequence of the
associated shuffle or quasi--shuffle (stuffle) algebras \cite{Hoffman}. These properties are 
widely independent of the specific realization of these algebras. To one of us (JB) it appeared
as a striking surprise, when finding the determinant-formula for harmonic sums of equal argument
\cite{Blumlein:1998if} Eqs.~(157,158) 
%----------------------------------------------------------------------------------------------
\begin{eqnarray}
S_{\underbrace{\mbox{\scriptsize a, \ldots ,a}}_{\mbox{\scriptsize $k$}}}(N)
= \frac{1}{k} \sum_{l=0}^k
S_{\underbrace{\mbox{\scriptsize a, \ldots ,a}}_{\mbox{\scriptsize $l$}}}(N)
S_{\wedge_{m=1}^{k-l} a}(N),~~
a \wedge b = {\rm sign}(a b)(|a|+|b|)
\end{eqnarray}
%----------------------------------------------------------------------------------------------
also in Ramanujan's notebook \cite{RAMA}, but for {\it integer sums}, which clearly differ in value from 
the former ones. Related relations to again different quantities were given by Fa\'a die 
Bruno \cite{FDBRUNO}. 

Iterated integrals with the same argument $x$ obey shuffle relations w.r.t. their product,
%----------------------------------------------------------------------------------------------
\begin{eqnarray}
\HA_{a_1,...,a_k}(x) \cdot \HA_{b_1,...,b_l}(x) = \sum_{\vec{c} \in \vec{a} \SH \vec{b}} 
\HA_{c_1,...c_{k+l}}(x).
\label{eq:SHUF} 
\end{eqnarray}
%----------------------------------------------------------------------------------------------
The shuffle-operation runs over all combinations of the sets $\vec{a}$ and $\vec{b}$ leaving
the order of these sets unchanged. Likewise, the (generalized) harmonic sums obey quasi-shuffle or 
stuffle-relations, which are found recursively using \cite{Moch:2001zr,Ablinger:2013cf}
%----------------------------------------------------------------------------------------------
\begin{eqnarray}   
        &&\S{a_1,\ldots ,a_k}{x_1,\ldots ,x_k;n}\S{b_1,\ldots ,b_l}{y_1,\ldots ,y_l;n}=\nonumber\\
        &&\hspace{2cm}\sum_{i=1}^n \frac{x_1^i}{i^{a_1}}\S{a_2,\ldots ,a_k}{x_2,\ldots ,x_k;i}
\S{b_1,\ldots ,b_l}{y_1,\ldots ,y_l,i} \nonumber\\
        &&\hspace{2cm}+\sum_{i=1}^n \frac{y_1^i}{i^{b_1}}
\S{a_1,\ldots ,a_k}{x_1,\ldots ,x_k,i}\S{b_2,\ldots ,b_l}{y_2,\ldots ,y_l;i} \nonumber\\
        &&\hspace{2cm}-\sum_{i=1}^n \frac{(x_1\cdot y_1)^i}{i^{a_1+b_1}}
\S{a_2,\ldots ,a_k}{x_2,\ldots ,x_k,i}\S{b_2,\ldots ,b_l}{y_2,\ldots ,y_l;i},\nonumber\\ &&
 \hspace{2cm} x_i, y_i \in \mathbb{C}, a_i, b_i \in \mathbb{N} \backslash \{0\}~.
        \label{SSsumproduct}
\end{eqnarray}
%-----------------------------------------------------------------------------------------
The presence of trace terms in form of lower weight products in addition to the shuffled 
terms, cf.~\cite{Blumlein:2003gb}, leads to the name stuffle relations. In case the corresponding
values exist, both (\ref{eq:SHUF},\ref{SSsumproduct}) can be applied to the multiple zeta values
or other special numbers applying the integral and sum-representations at  $x = 1$ and $N
\rightarrow \infty$, cf.~\cite{Borwein:1999js}.
The basis elements applying the (quasi) shuffle relations in case of the harmonic sums and 
polylogarithms at a given weight $w$ can be identified by the Lyndon words 
\index{Lyndon words}\cite{LYND,Reutenauer1993}. 
Let $\mathfrak{A} = \{a,b,c,d, ...\}$ be an {\it ordered} alphabet and $\mathfrak{A}^*(\mathfrak{A})$  
the set of words $w$ given as concatenation products. Under the ordering of $\mathfrak{A}$ a Lyndon word 
is smaller than any of its suffixes. For example, the set $\{a,a,a,b,b,b\},~a < b$ is associated to 
the Lyndon words $\{aaabbb, aababb, aabbab\}$. Radford showed \cite{RADF} that a shuffle algebra is
freely generated by the Lyndon words. The number of Lyndon words can be counted using 
Witt formulae \index{Witt formula}
\cite{WITT}. Let $\mathfrak{M}$ be a set of letters $q$ in which the letter $a_k$ emerges $n_k$
times, and $n = \sum_{k=1}^q n_k$. The number of Lyndon words associated to this set is given by 
%----------------------------------------------------------------------------------------------
\begin{eqnarray}   
l_n(n_1, \ldots, n_q) = \frac{1}{n} \sum_{d|n_k} \mu(d) \frac{(n/d)!}{(n_1/d)! ... (n_q/d)!}~.
\end{eqnarray}
%----------------------------------------------------------------------------------------------
Similarly one may count the basis elements occurring for all combinations at a given weight,
if the alphabet has $m$ letters~:
%----------------------------------------------------------------------------------------------
\begin{eqnarray}   
N_A(w) = \frac{1}{w} \sum_{d|w} \mu\left(\frac{w}{d}\right) m^d~,
\end{eqnarray}
%----------------------------------------------------------------------------------------------
where $\mu$ denotes M\"obius' function \cite{MOEBIUS}. In case of the harmonic sums and polylogarithms
one has $m=3$. The original number of harmonic polylogarithms is $3^w$
and in case of the harmonic sums $2 \cdot 3^{w-1}$. Algebraic relations for the harmonic 
polylogarithms and harmonic sums are implemented in the {\tt FORM}-codes {\tt summer} \cite{Vermaseren:1998uu}
and {\tt harmpol} \cite{Remiddi:1999ew}, {\tt HPL} \cite{Maitre:2005uu},
and also {\tt HarmonicSums.m} \cite{Ablinger:2013cf,JA1}.
%%%%%%%%%%%%%%%%%%%%%%%%%%%%%%%%%%%%%%%%%%%%%%%%%%%%%%%%%%%%%%%%%%%%%%%%%%%%%%%%%%%%%%%%%%%%%%%
\subsection{Structural Relations}
%%%%%%%%%%%%%%%%%%%%%%%%%%%%%%%%%%%%%%%%%%%%%%%%%%%%%%%%%%%%%%%%%%%%%%%%%%%%%%%%%%%%%%%%%%%%%%%

\vspace{1mm}\noindent
Structural relations of harmonic polylogarithms and harmonic sums are implied by
operations on their arguments $x$ and $N$, respectively.
%%%%%%%%%%%%%%%%%%%%%%%%%%%%%%%%%%%%%%%%%%%%%%%%%%%%%%%%%%%%%%%%%%%%%%%%%%%%%%%%%%%%%%%%%%%%%%%
\subsubsection{Harmonic Polylogarithms
\label{sec:HPLS}}
%%%%%%%%%%%%%%%%%%%%%%%%%%%%%%%%%%%%%%%%%%%%%%%%%%%%%%%%%%%%%%%%%%%%%%%%%%%%%%%%%%%%%%%%%%%%%%%
Harmonic polylogarithms satisfy argument-relations, as has been illustrated in (\ref{eq:SP1}--\ref{eq:SP2})
for some examples in case of the Nielsen integrals. Not all argument relations map inside
the harmonic polylogarithms, however, cf. \cite{Remiddi:1999ew}. Some of them are valid only 
for the sub-alphabet $\{\omega_0,\omega_1\}$. While the transformation $x \rightarrow -x$ is 
general
%----------------------------------------------------------------------------------------------
\begin{eqnarray}
\HA_{\vec{a}}(-x) = (-1)^p \HA_{-\vec{a}}(x), 
\end{eqnarray}
%-----------------------------------------------------------------------------------------   
with the last letter in $\vec{a}$ diffeent from 0 and $p$ the number of non-zero letters in $\vec{a}$. The transformations 
%----------------------------------------------------------------------------------------------
\begin{eqnarray}
x \rightarrow 1 - x, ~~~~~ x\rightarrow x^2
\end{eqnarray}
%-----------------------------------------------------------------------------------------   
apply to subsets only. Examples are~:
%----------------------------------------------------------------------------------------------
\begin{eqnarray}
\HA_{1, 0, 1}(1 - x) &=&
-\HA_0(x) \HA_{0,1}(x)+2 \HA_{0,0,1}(x)- \zeta_2 \HA_0(x)-2 \zeta_3
\\
  \HA_{1,0,0,1}\left(x^2\right) &=& 4 \left[
  \HA_{1,0,0,1}(x)
- \HA_{1,0,0,-1}(x)
- \HA_{-1,0,0,1}(x)
+ \HA_{-1,0,0,-1}(x)
\right].
\end{eqnarray}
%-----------------------------------------------------------------------------------------   
One may transform arguments by $x \rightarrow 1/y + i\varepsilon$,
%----------------------------------------------------------------------------------------------
\begin{eqnarray}
\HA_{1,0,1}\left(\frac{1}{x}\right) &=& 
\HA_0(x) \left[\HA_{0,1}(x)+i \pi  \HA_1(x)-4 \zeta_2 + \pi ^2\right]
-2 [\HA_{0,0,1}(x)-\HA_{0,1,1}(x) 
\nonumber\\ &&
+ \zeta_3]
+\left[-\HA_1(x)-i \pi \right] \HA_{0,1}(x) + 2 \zeta_2 \HA_1(x)-\frac{1}{6} \HA^3_0(x)
\nonumber\\ &&
+\frac{1}{2} i \left[\pi +i \HA_1(x)\right] \HA^2_0(x)~.
\end{eqnarray}
%-----------------------------------------------------------------------------------------   
An important general transformation is 
%----------------------------------------------------------------------------------------------
\begin{eqnarray}
\label{eq:conf}
x \rightarrow \frac{1-t}{1+t}
\end{eqnarray}
%-----------------------------------------------------------------------------------------   
which acts on the HPLs but not on the subset of Nielsen-integrals. 
An example is~:
%----------------------------------------------------------------------------------------------
\begin{eqnarray}
\HA_{1, -1, 0}\left(\frac{1 - x}{1 + x}\right) &=&
 \frac{1}{6} \HA^3_{-1}(x)
+ \HA_{-1,-1,1}(x)
- \HA_{0,-1,-1}(x)
- \HA_{0,-1,1}(x)
+ \frac{15 \zeta_3}{8}
\nonumber\\ &&
-\frac{1}{2} \zeta_2 \left[\HA_{-1}(x)-\HA_0(x)\right]
-2 \left[\frac{\zeta_3}{8}-\frac{\ln(2) \zeta_2}{2}\right]
-2 \ln(2) \zeta_2~. 
\nonumber\\
\end{eqnarray}
%-----------------------------------------------------------------------------------------   
In most of these relations also HPLs at argument $x=1$ contribute, cf.~Sect.~\ref{sec:5}.
Structural relations of HPLs are implemented in the packages 
{\tt harmpol} \cite{Remiddi:1999ew}, {\tt HPL} \cite{Maitre:2005uu},
and {\tt HarmonicSums.m} \cite{Ablinger:2013cf,JA1}.
%%%%%%%%%%%%%%%%%%%%%%%%%%%%%%%%%%%%%%%%%%%%%%%%%%%%%%%%%%%%%%%%%%%%%%%%%%%%%%%%%%%%%%%%%%%%%%%
\subsubsection{Harmonic Sums}
%%%%%%%%%%%%%%%%%%%%%%%%%%%%%%%%%%%%%%%%%%%%%%%%%%%%%%%%%%%%%%%%%%%%%%%%%%%%%%%%%%%%%%%%%%%%%%%
Harmonic sums obey the duplication relation
%----------------------------------------------------------------------------------------------
\begin{eqnarray}   
S_{i_1,...,i_n}(N) = 2^{i_1+...+i_n-n} \sum_\pm S_{\pm i_1,...,\pm i_n}(2N),~~i_k \in \mathbb{N} 
\backslash \{0\}~.
\label{eq:DUP}
\end{eqnarray}
%-----------------------------------------------------------------------------------------
This allows to define harmonic sums at half-integer, i.e. rational, values. Ultimately, one would 
like to derive expressions for $N \in \mathbb{C}$, cf.~Sect.~\ref{sec:9}. Another extension is to
$N \in \mathbb{R}$ \cite{Blumlein:1998if,Blumlein:2009ta}. The representation of harmonic sums
through Mellin-transforms (\ref{eq:MEL}) implies analyticity for a finite range around a given value of
$N$. The Mellin-transform of a harmonic polylogarithm can thus be differentiated for $N$
%----------------------------------------------------------------------------------------------
\begin{eqnarray}   
\label{eq:DIFF}
\frac{d}{dN} \int_0^1 dx x^{N-1} \HA_{\vec{a}}(x) = \int_0^1 dx x^{N-1} \HA_0(x) \HA_{\vec{a}}(x)~.
\end{eqnarray}
%-----------------------------------------------------------------------------------------
In turn, the shuffling relation (\ref{eq:SHUF}) allows to represent the r.h.s. in (\ref{eq:DIFF})
as the Mellin-transform of other HPLs. It turns out that differentiation of harmonic sums for $N$
is closed under additional association of the multiple zeta values \cite{Blumlein:2009ta}.
The number of basis elements by applying the duplication relation (H), resp. its combination with 
the algebraic relations is \cite{ABS13}
%----------------------------------------------------------------------------------------------
\begin{eqnarray}   
N_H(w) = 2 \cdot 3^{w-1} - 2^{w-1},~~~~~N_{AH}(w) =  \frac{1}{w} \sum_{d|w} \mu\left(\frac{w}{d}\right) 
\left[2^2 - 3^d\right]~.
\end{eqnarray}
%-----------------------------------------------------------------------------------------
Differentiation in combination with the other relation yields
%----------------------------------------------------------------------------------------------
\begin{eqnarray}   
N_D(w) = 4 \cdot 3^{w-1}, N_{DH}(w) = 4 \cdot 3^{w-2} - 2^{w-2}, 
N_{ADH}(w) = N_{AH}(w) - N_{AH}(w-1). \nonumber\\
\end{eqnarray}
%-----------------------------------------------------------------------------------------

Let us close with a remark on observables or related quantities in physics which are 
calculated to a certain loop level and can be thoroughly expressed in terms of harmonic sums. 
As a detailed investigation of massless and single mass 2-loop quantities showed \cite{TWOLOOP} 
seven basic functions of up to weight $w=4$, cf. \cite{Blumlein:2009ta}, are sufficient 
to express all quantities. The 3-loop anomalous dimensions \cite{ANDI3} contributing to the $1/\varepsilon$ 
poles of the corresponding matrix elements require fifteen functions of up to  weight $w=5$ 
and further twenty basic functions are needed
to also express the massless Wilson coefficients \cite{Vermaseren:2005qc} in deep-inelastic 
scattering \cite{Blumlein:2012bf}, cf.~Ref.~\cite{Blumlein:2009fz}. Despite of the complexity of these 
calculations finally a rather compact structure is obtained for the representation of the results.
Structural relations of harmonic sums are implemented in the package
{\tt HarmonicSums.m} \cite{Ablinger:2013cf,JA1}.
%%%%%%%%%%%%%%%%%%%%%%%%%%%%%%%%%%%%%%%%%%%%%%%%%%%%%%%%%%%%%%%%%%%%%%%%%%%%%%%%%%%%%%%%%%%%%%%
\section{Multiple Zeta Values}
\label{sec:5}
%%%%%%%%%%%%%%%%%%%%%%%%%%%%%%%%%%%%%%%%%%%%%%%%%%%%%%%%%%%%%%%%%%%%%%%%%%%%%%%%%%%%%%%%%%%%%%%

\vspace{1mm}\noindent
The multiple zeta values (MZVs) \cite{EZ,Zagier1994}\footnote{For a detailed account on the 
literature on MZVs see \cite{Blumlein:2009cf,HOFF1} and the surveys Ref.~\cite{SURVMZV}.} 
are obtained by the limit $N \rightarrow \infty$ of the harmonic sums 
%----------------------------------------------------------------------------------------------
\begin{eqnarray}   
\lim_{N\rightarrow \infty} S_{\vec{a}}(N) = \sigma_{\vec{a}} 
\end{eqnarray}
%-----------------------------------------------------------------------------------------
and may also be represented 
in terms of linear combinations of harmonic polylogarithms $\HA_{\vec{b}}(1)$ over the alphabet
$\{\omega_0, \omega_1, \omega_{-1}\}$.\footnote{The numbers associated with this alphabet are 
sometimes also called Euler-Zagier values and those of the sub-alphabet $\{\omega_0,\omega_1\}$ 
multiple zeta values.} In the former case one usually includes the divergent 
harmonic sums since all divergent contributions are uniquely represented in terms of polynomials 
in $\sigma_1(\infty) \equiv \sigma_0$ due to the algebraic relations. Likewise, not all harmonic 
polylogarithms can be calculated at $x=1$, requiring their re-definition in terms of distributions. 
Some examples for MZVs, which already appear in case of Nielsen
integrals, are~:  
%---------------------------------------------------------------------------------------------
\begin{eqnarray}
\label{eq:a3}
\Li_n(1)   &=& \zeta_n \\
\Li_n(-1)  &=& -\left(1-\frac{1}{2^{n-1}}\right)\zeta_n \\
{\sf S}_{1,p}(1) &=& \zeta_{p+1}\\
\label{eq:a4}
{\sf S}_{1,2}(-1) &=& \frac{1}{8}\zeta_{3}\\
\Li_2\left(\frac{1}{2}\right) &=& \frac{1}{2}\left[\zeta_2 - \ln^2(2)\right]\\
\Li_3\left(\frac{1}{2}\right) &=& \frac{7}{8} \zeta_3 - \frac{1}{2} \zeta_2 \ln(2) + \frac{1}{6} 
\ln^3(2) \\
{\sf S}_{2,2}(1)  &=& \frac{1}{10} \zeta_2^2\\ 
\label{eq:a5}
{\sf S}_{2,2}(-1) &=& - \frac{3}{4} \zeta_2^2 + 2 \Li_4\left(\frac{1}{2}\right) + \frac{7}{4} \zeta_3 
\ln(2) - \frac{1}{2} \zeta_2 \ln^2(2) + \frac{1}{12} \ln^4(2) 
\\
\label{eq:a6}
{\sf S}_{1,3}(-1) &=& - \frac{2}{5} \zeta_2^2 + \Li_4\left(\frac{1}{2}\right) + \frac{7}{8} \zeta_3   
\ln(2) - \frac{1}{4} \zeta_2 \ln^2(2) + \frac{1}{24} \ln^4(2)~.
\end{eqnarray}
%---------------------------------------------------------------------------------------------
In case of physics applications, MZVs played a role in loop calculations rather early, 
cf. \cite{PHYS}. Since for $\Li_m(1/2)$ for $m = 2,3$ these numbers are not elementary, 
(\ref{eq:a5}, \ref{eq:a6}) seem to fail to provide a corresponding relation
for $m = 4$. Similarly, for larger values of $m$ also no reduction has been observed.

A central question concerns the representation of harmonic sums in terms of polynomial bases.
This has been analyzed systematically in \cite{Blumlein:2009cf,KV}.~\footnote{For some aspects of the earlier
development including results by the Leuven-group, Zagier, Broadhurst, Vermaseren and the Lille-group, 
see \cite{Blumlein:2009cf}.} For MZVs over $\{0,1\}$ a proof on the maximum of basis elements 
at fixed weight $w$ has been given in Refs.~\cite{W1}. At the lowest weights the shuffle and 
stuffle relations imply all relations for the MZVs. Starting with weight $w = 8$ one also needs
the duplication relation (\ref{eq:DUP}), and from weight $w = 12$ also the generalized 
duplication relations Sect.~4.1 in \cite{Blumlein:2009cf}. The latter are closely related to the conformal 
transformation relations of the HPLs at $x=1$, see (\ref{eq:conf}). Let us give one example for the
combined use of the shuffle and stuffle relation for illustration, \cite{Borwein:1999js}\footnote{
Here the $\zeta_{\vec{a}}$-values are defined $\zeta_{a_1,...,a_m} = \sum_{n_1 > n_2 > ...> n_m}^\infty
\prod_{k=1}^m n_k^{-a_1}$}.~:
%---------------------------------------------------------------------------------------------
\begin{eqnarray}
\text{shuffle}~:~~\zeta_{2,1} \zeta_2 &=& 6 \zeta_{3,1,1} + 2 \zeta_{2,2,1} + \zeta_{2,1,2}
\nonumber\\ 
\text{stuffle}~:~~\zeta_{2,1} \zeta_2 &=& 2 \zeta_{2,2,1} + \zeta_{41} + \zeta_{2,3} + \zeta_{2,1,2}
\nonumber\\
\Longrightarrow~~~\zeta_{3,1,1} &=& \frac{1}{6} \left[ \zeta_{4,1} + \zeta_{2,3} - \zeta_{2,2,1}\right]~.
\end{eqnarray} 
%---------------------------------------------------------------------------------------------
Finally one derives a basis for the MZVs using the above relations.
Up to weight $w = 7$ reads, cf.~\cite{Vermaseren:1998uu},
%---------------------------------------------------------------------------------------------
\begin{eqnarray}
&& \Biggl\{
\left(\sigma_1(\infty), \ln(2)\right); 
\zeta_2; 
\zeta_3; 
\Li_4\left(\tfrac{1}{2}\right);
\left(\zeta_5,\Li_5\left(\tfrac{1}{2}\right)\right);
\left(\Li_6\left(\tfrac{1}{2}\right),\sigma_{-5,-1}\right);
\nonumber\\ &&
\left(\zeta_7,\Li_7\left(\tfrac{1}{2}\right),\sigma_{-5,1,1},\sigma_{-5,1,1}\right);\Biggr\}
\end{eqnarray}
%---------------------------------------------------------------------------------------------
In \cite{Blumlein:2009cf} bases were calculated up to $w=12$ for the alphabet $\{0,1,-1\}$ and
to $w = 22$ for the alphabet $\{0,1\}$ in explicit form resp. for $w = 24$ restricting to basis
elements only. In the latter case the conjecture by Zagier \cite{Zagier1994} that the shuffle, stuffle 
and duplication relations were the only ones was confirmed up to the weights quoted. 
For these cases counting relations were conjectured in Refs.~\cite{Broadhurst:1996az,Broadhurst:1996kc}.
One may represent the basis
for the MZVs over the alphabets $\{0,1\}$ resp. $\{0,1,-1\}$ by polynomial bases or count just
the factors appearing in these polynomials of special numbers occurring newly in the corresponding weight, which is 
called Lyndon-basis \cite{Blumlein:2009cf}.
In the first case the basis for the MZVs[$\{0,1\}$] is conjectured to be counted by the 
Padovan numbers \index{Padovan numbers}
$\hat{P}_k$ \cite{PADO} generated by
%---------------------------------------------------------------------------------------------
\begin{eqnarray}
\frac{1+x}{1 - x^2 - x^3} = \sum_{k=0}^\infty x^k \hat{P}_k,~~~\hat{P}_1 = \hat{P}_2 = \hat{P}_3 = 1.
\end{eqnarray}
%---------------------------------------------------------------------------------------------
In case of the Lyndon basis the Perrin numbers \index{Perin numbers} $P_k$ appear 
\cite{PERRIN}
%---------------------------------------------------------------------------------------------
\begin{eqnarray}
\frac{3-x^2}{1 - x^2 - x^3} &=& \sum_{k=0}^\infty x^k \hat{P}_k,~~~P_1 = 0, P_2 = 2, P_3, =3~.
\end{eqnarray}
%---------------------------------------------------------------------------------------------
Both the above sequences obey the Fibonacci-recurrence\index{Fibonacci numbers} 
\cite{FIBO}
%---------------------------------------------------------------------------------------------
\begin{eqnarray}
P_d = P_{d-2} + P_{d-3},~~~ d \geq 3~.
\end{eqnarray}
%---------------------------------------------------------------------------------------------
The length of the Lyndon basis at weight $w$ is given by
%---------------------------------------------------------------------------------------------
\begin{eqnarray}
l(w) &=& \frac{1}{w} \sum_{d|w} \mu\left(\frac{w}{d}\right) P_d~.
\end{eqnarray}
%---------------------------------------------------------------------------------------------
Hoffman \cite{HOFFC} conjectured that all MZVs over the alphabet $\{0,1\}$ can be represented
over a basis of MZVs carrying 2 and 3 as indices only. This has been confirmed up to $w = 24$.
An explicit proof has been given in \cite{BROWNH}.

The polynomial basis of the MZVs[$\{0,1,-1\}$] is conjectured to be counted by the Fibonacci numbers 
\cite{BINET}
%---------------------------------------------------------------------------------------------
\begin{eqnarray}
f_d = \frac{1}{\sqrt{5}}\left[\left(\frac{1+\sqrt{5}}{2}\right)^d - \left(\frac{1-\sqrt{5}}{2}\right)^d
\right]
\end{eqnarray}
%---------------------------------------------------------------------------------------------
which obey
%---------------------------------------------------------------------------------------------
\begin{eqnarray}
\frac{x}{1-x-x^2} = \sum_{k=0}^\infty x^k f_k~.
\end{eqnarray}
%---------------------------------------------------------------------------------------------
For the corresponding Lyndon basis the counting relation
%---------------------------------------------------------------------------------------------
\begin{eqnarray}
l(w) &=& \frac{1}{w} \sum_{d|w} \mu\left(\frac{w}{d}\right) L_d
\end{eqnarray}
%---------------------------------------------------------------------------------------------
is conjectured \cite{Broadhurst:1996az,Broadhurst:1996kc}, where $L_d$ are the 
Lucas numbers \index{Lucas numbers}\cite{LUCAS}, 
%---------------------------------------------------------------------------------------------
\begin{eqnarray}
L_d &=& \left(\frac{1+\sqrt{5}}{2}\right)^d + \left(\frac{1-\sqrt{5}}{2}\right)^d \\
\frac{2-x}{1-x-x^2} &=& \sum_{k=0}^\infty x^k L_k, \\
& & L_d = L_{d-1} + L_{d-2}, d \geq 4,~~~~L_1 = 1, L_2 = 3, L_3 = 4.
\end{eqnarray}
%---------------------------------------------------------------------------------------------

There is a series of Theorems proven on the MZVs, see also \cite{Blumlein:2009cf}, which can be verified 
using the data base \cite{Blumlein:2009cf}. The {\sf duality theorem} \index{duality 
theorem}
~\cite{Zagier1994} in case of the alphabet 
$\{\omega_0,\omega_1\}$ states
%---------------------------------------------------------------------------------------------
\begin{eqnarray}
\label{eq:dual}
\HA_{\vec{a}}(1) =
\HA_{\vec{a}^\dagger}(1),~~~\vec{a}^\dagger = \vec{a}^T_{0 \longleftrightarrow 1}~.
\end{eqnarray} 
%---------------------------------------------------------------------------------------------
In case of the alphabet $\{\omega_0,\omega_1,\omega_{-1}\}$ it is implied by the transformation
(\ref{eq:conf}), see \cite{Blumlein:2009cf}. 
Another relation is the {\sf sum theorem},\index{sum theorem}~Ref.~\cite{Euler,SUMTH},
%-------------------------------------------------------------------------------
\begin{equation}
\sum_{i_1 + \ldots +i_k =n, i_1 > 1} \zeta_{i_1, \ldots, i_k} = \zeta_n~.
\end{equation}
%-------------------------------------------------------------------------------
The sum-theorem was conjectured in \cite{HOF3}, cf.~\cite{HOF2}. For its derivation
using the Euler connection formula for polylogarithms, cf.~\cite{OU}.

Further identities are given by the {\sf derivation theorem}, \index{derivation 
theorem} \cite{HOF3,HOF4}.
Let $I=(i_1, \ldots, i_k)$ any sequence of positive integers with $i_1 > 1$.
Its derivation $D(I)$ is given by
%-------------------------------------------------------------------------------
\begin{eqnarray}
D(I) &=&
        (i_1 + 1, i_2, \ldots, i_k) +
(i_1,   i_2+1, \ldots, i_k) + \ldots
(i_1,   i_2, \ldots, i_k+1) \nonumber\\
\zeta_{D(I)} &=& \zeta_{(i_1 + 1, i_2, \ldots, i_k)} + \ldots
+ \zeta_{(i_1,   i_2, \ldots, i_k+1)}~.
\end{eqnarray}
%-------------------------------------------------------------------------------   
The derivation theorem states
%-------------------------------------------------------------------------------
\begin{equation}
\zeta_{D(I)} = \zeta_{\tau(D(\tau(I)))}~,
\end{equation}
%-------------------------------------------------------------------------------
where $\tau$ denotes the duality-operation, cf.~(\ref{eq:dual}).
An index-word $w$ is called admissible, if its first letter is not {\sf 1}. The words form the set
$\mathfrak{H}^0$. $|w|$ = {\sf w} is the weight and $d(w)$ the depth of $w$.
For the MZVs the words $w$ are build in terms of concatenation
products $x_0^{i_1 - 1} x_1 x_0^{i_2 - 1} x_0 ... x_0^{i_k - 1} x_1$.
The height of a word, ht$(w)$, counts the number of (non-commutative) factors
$x_0^a x_1^b$ of $w$. The operator $D$ and
its dual $\overline{D}$ act as follows \cite{ZUD},
%-------------------------------------------------------------------------------
\begin{equation}
D x_0 =0,~~~~ D x_1 = x_0 x_1,~~~~ \overline{D} x_0 = x_0 x_1,~~~~
\overline{D} x_1 = 0~.
\nonumber
\end{equation}
%-------------------------------------------------------------------------------
Define an anti-symmetric derivation
%-------------------------------------------------------------------------------
\begin{equation}
\partial_n x_0 = x_0 (x_0 + x_1)^{n-1} x_1~.
\nonumber
\end{equation}
%-------------------------------------------------------------------------------
A generalization of the derivation theorem was given in
\cite{HOF4,KANEK1}. The identity
%-------------------------------------------------------------------------------
\begin{equation}
\zeta(\partial_n w) = 0
\end{equation}
%-------------------------------------------------------------------------------
holds for any $n \geq 1$ and any word $w \in \mathfrak{H}^0$.
Further theorems are the  {\sf Le--Murakami theorem},~\cite{LEMURA},
the {\sf Ohno theorem},~\cite{OHNO}, which generalizes the sum- and
duality theorem, the {\sf Ohno--Zagier theorem}, \cite{OHNZAG},
which covers the Le--Murakami theorem and the sum theorem,
which generalizes a theorem by Hoffman~\cite{HOF3,HOF2}, and the {\sf cyclic sum
theorem},~\cite{OHNO1}.

There are also relations for MZVs at repeated arguments\index{Zeta values at repeated 
arguments}, cf.~\cite{Borwein:1999js,ZUD,Borwein:1996yq},
on which examples are~:
%---------------------------------------------------------------------------------------------
\begin{eqnarray}
\zeta(\{2\}_n)  &=& \frac{2(2\pi)^{2n}}{(2n+1)!} \frac{1}{2}
\\
\zeta(2,\{1\}_n) &=& \zeta(n+2)
\\
\zeta(\{3,1\}_n) &=& \frac{1}{4^n} \zeta(\{4\}_n) = \frac{2 \pi^{4n}}{(4n+2)!}
\\
\zeta(\{10\}_n) &=& \frac{10 (2\pi)^{2n}}{(10n+5)!} \left[1
+\left(\frac{1+\sqrt{5}}{2}\right)^{10n+5}
+\left(\frac{1-\sqrt{5}}{2}\right)^{10n+5} \right]~.
\end{eqnarray}
%---------------------------------------------------------------------------------------------

Finally, we mention a main conjecture for the MZVs over $\{0,1\}$. Consider tuples ${\bf k} 
= (k_1, \ldots, k_r)~\in~\mathbb{N}^r, k_1 \geq 1$. One defines
%-------------------------------------------------------------------------------------------
\begin{eqnarray}
{\mathcal Z}_0 &:=& \mathbb{Q} \nonumber\\
{\mathcal Z}_1 &:=& \{0\} \nonumber\\
{\mathcal Z}_w &:=& \sum_{|{\bf k}| = w} \mathbb{Q}
 \cdot \zeta({\bf k}) \subset \mathbb{R}~.
\end{eqnarray}
%-------------------------------------------------------------------------------------------
If further
%-------------------------------------------------------------------------------------------
\begin{eqnarray}
{\mathcal Z}^{\rm Go} &:=& \sum_{w=0}^\infty {\mathcal Z}_w \subset
\mathbb{R}~~~~~~{\rm (Goncharov)} \\
{\mathcal Z}^{\rm Ca} &:=& \overset{\infty}{\underset{w=0}{
\bigoplus}} {\mathcal Z}_w ~~~~~~~~~~~~~~~{\rm (Cartier)}
\end{eqnarray}
%-------------------------------------------------------------------------------------------
the conjecture states

\vspace*{1mm} \noindent
(a)~~${\mathcal Z}^{\rm Go} \cong {\mathcal Z}^{\rm Ca}$. There are no
relations over $\mathbb{Q}$ between the MZVs of different weight $w$.\\
(b)~~dim${\mathcal Z}_w  = d_w$, with $d_0 = 1, d_1=0, d_2=1, d_w = d_{w-2}
+d_{w-3}$.\\
(c)~~All algebraic relations between MZVs are given by the extended double-shuffle
relations~\cite{IKZ}, cf. also \cite{ECALLE}. If this conjecture turns out to
be true all MZVs are irrational numbers.

Let us also mention a few interesting relations for $\Li_2(z)$ for special arguments 
found by Ramanujan \cite{RAMANUJAN}, which are of use e.g. in massive calculations at 
3-loops \cite{ABRSW13}. These numbers relate to constants beyond the MZVs, which occur for generalized sums
and their extension allowing for binomial and inverse binomial weights~:
%---------------------------------------------------------------------------------------------
\begin{eqnarray}
\Li_2\left(\frac{1}{3}\right) - \Li_2\left(\frac{1}{9}\right) 
&=& \frac{\pi^2}{18} - \frac{1}{6}\ln^2(3)
\\
\Li_2\left(-\frac{1}{2}\right) + \frac{1}{6} \Li_2\left(\frac{1}{9}\right)
&=& - \frac{\pi^2}{18} + \ln(2) \ln(3) - \frac{1}{2} \ln^2(2) - \frac{1}{3} \ln^2(3)
\\
\Li_2\left(\frac{1}{4}\right) + \frac{1}{3} \Li_2\left(\frac{1}{9}\right)
&=&  \frac{\pi^2}{18} + 2 \ln(2) \ln(3) - 2 \ln^2(2) - \frac{2}{3} \ln^2(3)
\\
\Li_2\left(-\frac{1}{3}\right) - \frac{1}{3}\Li_2\left(\frac{1}{9}\right)
&=& - \frac{\pi^2}{18} + \frac{1}{6} \ln^2(3)
\\
\Li_2\left(-\frac{1}{8}\right) + \frac{1}{3}\Li_2\left(\frac{1}{9}\right)
&=& - \frac{1}{2} \ln^2\left(\frac{9}{8}\right)~. 
\end{eqnarray}
%---------------------------------------------------------------------------------------------
For further specific numbers, which occur in the context of Quantum Field Theory calculations 
see also Sections~\ref{sec:Snum}, \ref{sec:7}, and \ref{sec:9}.
%%%%%%%%%%%%%%%%%%%%%%%%%%%%%%%%%%%%%%%%%%%%%%%%%%%%%%%%%%%%%%%%%%%%%%%%%%%%%%%%%%%%%%%%%%%%%%%
\section{Generalized Harmonic Sums and Polylogarithms}
\label{sec:6}
%%%%%%%%%%%%%%%%%%%%%%%%%%%%%%%%%%%%%%%%%%%%%%%%%%%%%%%%%%%%%%%%%%%%%%%%%%%%%%%%%%%%%%%%%%%%%%%
\subsection{Generalized Harmonic Sums}
%%%%%%%%%%%%%%%%%%%%%%%%%%%%%%%%%%%%%%%%%%%%%%%%%%%%%%%%%%%%%%%%%%%%%%%%%%%%%%%%%%%%%%%%%%%%%%%

\vspace{1mm}\noindent
Generalized harmonic sums\index{Generalized harmonic sums}, also called S--sums\index{S-sums}, are 
defined by 
\cite{Moch:2001zr,Ablinger:2013cf,Weinzierl:2013yn}
%---------------------------------------------------------------------------------------------
\begin{eqnarray}
S_{a_1, \ldots, a_k}(x_1, \ldots, x_k;N) &=& \sum_{i_1=1}^N \frac{x_1^{i_1}}{i_1^{a_1}} S_{a_2, 
\ldots, a_k}(x_2, \ldots, x_k;i_1), \nonumber\\ && 
~~S_\emptyset = 1~,~~x_i \in \mathbb{R} \backslash \{0\},~~a_i \in \mathbb{N} \backslash \{0\}
\end{eqnarray}
%---------------------------------------------------------------------------------------------
and form a quasi-shuffle and a Hopf algebra\index{Hopf algebra} \cite{HOPF} under the 
multiplication (\ref{SSsumproduct}) \cite{Moch:2001zr}. 
The multiplication relation in general leads outside the weight sets $\{a_i\}, \{b_i\}$.
The S--sums cover (together with the limit $N \rightarrow \infty$) the classical polylogarithms, the Nielsen 
functions, the harmonic 
polylogarithms, the multiple polylogarithms\index{multiple polylogarithm} 
\index{multiple polylogarithm} \cite{GON1}, 
the two-dimensional HPLs \index{two-dimensional harmonic polylpgarithm} \cite{TDHPL},
and the MZVs 
\cite{Moch:2001zr}. In Ref.~\cite{Moch:2001zr} four algorithms were presented allowing to 
perform the $\varepsilon$--expansion of classes of sums in terms of S--sums, which were coded in
two packages \cite{Weinzierl:2002hv,Moch:2005uc}. In this way the $\varepsilon$--expansion can be 
performed using convergent serial representations  for the generalized hypergeometric 
functions \index{generalized hypergeometric functiuons}
$_PF_Q$, The Appell-functions $F_{1,2}$ \index{Appell functions}, and the Kamp\'{e} de 
F\'{e}riet function \index{Kamp\'e de F\'eriet functions}\cite{HTR}.

They can be represented in terms of a Mellin transformation over $x \in [0,x_1...x_k]$ 
\cite{Ablinger:2013cf}. E.g.~the single sums are given by
%-----------------------------------------------------------------------------------------
\begin{eqnarray}   
{\rm S}_m(b;N) = \int_0^b \frac{dx_m}{x_m} ... \int_0^{x_3} \frac{dx_2}{x_2} \int_0^{x_2} 
d x_1 \frac{x_1^N-1}{x_1 -1}~.
\end{eqnarray}
%-----------------------------------------------------------------------------------------
Generalized harmonic sums obey the duplication relation
%-----------------------------------------------------------------------------------------
\begin{eqnarray}   
\sum{\S{a_m,\ldots,a_1}{\pm b_m,\ldots,\pm 
b_1;2\;N}}=\frac{1}{2^{\sum_{i=1}^m{a_i}-m}}\S{a_m,\ldots,a_1}{b_m^2,\ldots,b_1^2;N},
\end{eqnarray}
%-----------------------------------------------------------------------------------------
where the  sum on the left hand side is over the $2^m$ possible combinations concerning $\pm$ and
$a_i \in \N$, $b_i \in \R \backslash \{0\}$ and $n\in \N$.
They also obey differential relations w.r.t. $N$, supplementing their set with the generalized 
harmonic sums at infinity, resp. of the generalized harmonic polylogarithms at $x = 1$. The 
mapping will usually also require objects with different weights $x_i$. Examples are
\cite{Ablinger:2013cf}~:
%-----------------------------------------------------------------------------------------
\begin{eqnarray}   
\frac{\partial}{\partial n}\S{2}{2;n} &=& 
-\S{3}{2;n}+\H{0}2\S{2}{2;n}+\H{0,0,-1}1+2\H{0,0,1}1+\H{0,1,-1}1,
\nonumber\\ 
\frac{\partial}{\partial N}\textnormal{S}_3\hspace{-0.2em}\left(\tfrac{1}{4};N\right)
&=& 12 \Biggl[
-\textnormal{S}_{3,1}\hspace{-0.2em}\left(\tfrac{1}{2},\tfrac{1}{2};N\right)
-\frac{1}{2}
   \frac{\partial}{\partial N}\textnormal{S}_{2,1}\hspace{-0.2em}
\left(\tfrac{1}{2},\tfrac{1}{2};N\right)-\frac{1}{2} \textnormal{H}_{1,0}
\hspace{-0.2em}\left(\tfrac{1}{2}\right) \textnormal{S}_2\hspace{-0.2em}
\left(\tfrac{1}{2};N\right)\nonumber\\ &&
+\textnormal{H}_0\hspace{-0.2em}
\left(\tfrac{1}{2}\right) \textnormal{S}_{2,1}\hspace{-0.2em}
\left(\tfrac{1}{2},\tfrac{1}{2};N\right)-\frac{1}{2}
   \textnormal{H}_{\tfrac{1}{2}}\hspace{-0.2em}\left(\tfrac{1}{4}\right) 
\textnormal{H}_{0,1,0}\hspace{-0.2em}\left(\tfrac{1}{2}\right)+\frac{1}{12} 
\textnormal{H}_{0,0,1,0}\hspace{-0.2em}\left(\tfrac{1}{4}\right)\nonumber\\
\nonumber\\ && +\frac{1}{2} \textnormal{H}_{\tfrac{1}{2},0,1,0}\hspace{-0.2em}
\left(\tfrac{1}{4}\right)-\frac{1}{12} \textnormal{H}_0\hspace{-0.2em}
\left(\tfrac{1}{4}\right) \textnormal{S}_3\hspace{-0.2em}\left(\tfrac{1}{4};N\right)-\frac{1}{4}
   \textnormal{S}_2\hspace{-0.2em}\left(\tfrac{1}{2};N\right)^2 \Biggr]~.
\end{eqnarray}
%-------------------------------------------------------------------------------------
The counting relations for the basis elements are 
%-----------------------------------------------------------------------------------------
\begin{eqnarray}
N_D(w) = N_S(w) - N_S(w-1),~~~~~~N_{A,D}(w) = N_A(w) - N_A(w-1),
\end{eqnarray}
%-----------------------------------------------------------------------------------------
where $N_S = (n-1) \cdot n^{w-1}$ denotes the number of sums, given $n$ letters in the alphabet, and $N_A$ the 
basis elements after applying the algebraic 
equations. Explicit bases for a series of alphabets have been calculated in 
\cite{Ablinger:2013cf}.
%%%%%%%%%%%%%%%%%%%%%%%%%%%%%%%%%%%%%%%%%%%%%%%%%%%%%%%%%%%%%%%%%%%%%%%%%%%%%%%%%%%%%%%%%%%%%%%
\subsection{Generalized Harmonic Polylogarithms}
%%%%%%%%%%%%%%%%%%%%%%%%%%%%%%%%%%%%%%%%%%%%%%%%%%%%%%%%%%%%%%%%%%%%%%%%%%%%%%%%%%%%%%%%%%%%%%%

\vspace{1mm} \noindent
Generalized harmonic polylogarithms are defined as the Poincar\'{e}-iterated integrals 
\cite{POINC,POLYLOG2}
%-----------------------------------------------------------------------------------------
\begin{eqnarray}
\label{eq:GHPL}
\HA_{\vec a}(x) = \int_0^x \prod_{j=1}^m \frac{dz_j}{\abs{a_j} - \sign{a_j}z_j},~~~a_j \in \mathbb{C},~~~
z_j \geq z_{j+1}~.
\end{eqnarray}
%-----------------------------------------------------------------------------------------
For $a_j \in \mathbb{R}, a_j < 1$ (\ref{eq:GHPL}) is defined as Cauchy principal value only.
Already A.~Jonqui\`ere \cite{POLYLOG2} has studied integrals of this type. Sometimes they are
also called Chen-iterated integrals\index{Chen iterated integral}, cf. \cite{POINC}, or 
Goncharov polylogarithms \index{Goncharov polylogarithm}\cite{GON1}. 

The Mellin transforms of generalized harmonic polylogarithms map onto generalized harmonic sums
\cite{Ablinger:2013cf}. Furthermore, the generalized harmonic polylogarithms obey various argument 
relations similar to the case if the HPLs, cf. Sect.~\ref{sec:HPLS}, as
%-----------------------------------------------------------------------------------------
\begin{eqnarray}
x + b           &\rightarrow&  x \\
b - x           &\rightarrow&  x \\ 
\frac{1-x}{1+x} &\rightarrow& x  \\
             kx &\rightarrow& x  \\
\frac{1}{x}     &\rightarrow& x~.
\end{eqnarray}
%-----------------------------------------------------------------------------------------
%%%%%%%%%%%%%%%%%%%%%%%%%%%%%%%%%%%%%%%%%%%%%%%%%%%%%%%%%%%%%%%%%%%%%%%%%%%%%%%%%%%%%%%%%%%%%%%
\subsection{Relations between $S$-Sums at Infinity}
\label{sec:Snum}
%%%%%%%%%%%%%%%%%%%%%%%%%%%%%%%%%%%%%%%%%%%%%%%%%%%%%%%%%%%%%%%%%%%%%%%%%%%%%%%%%%%%%%%%%%%%%%%

\vspace{1mm} \noindent
S-sums at infinity exhibit a more divergent behaviour than harmonic sums if $a_1 > 1$. The degree 
of divergence is then at least $\propto a_1^N$, cf. Sect.~\ref{sec:9}. In the following we will 
discuss only convergent S-sums at infinity. They obey stuffle and shuffle relations, the 
duplication relation $N \rightarrow 2N$, and the duality relations for the 
generalized polylogarithms \cite{Ablinger:2013cf}
%----------------------------------------------------------------------------------------------
\begin{eqnarray}
\label{eq:du1}
1 - x             &\rightarrow& x \\
\label{eq:du2}
\tfrac{1-x}{1+x}  &\rightarrow& x \\
\label{eq:du3}
\tfrac{c-x}{d+x}  &\rightarrow& x,~~ c,d \in \R, d \neq 1~.
\end{eqnarray}
%----------------------------------------------------------------------------------------------
Eq.~(\ref{eq:du1}) implies
%----------------------------------------------------------------------------------------------
\begin{eqnarray}
\HA_{a_1,...,a_k}(1) = 
\HA_{1-a_1,...,1-a_k}(1), ~~a_k \neq 0~.
\end{eqnarray}
%----------------------------------------------------------------------------------------------
Examples for (\ref{eq:du2}, \ref{eq:du3}) are~:
%----------------------------------------------------------------------------------------------
\begin{eqnarray}
S_1\left(\tfrac{1}{2}; \infty\right) &=& - S_{-1}(\infty) \equiv \ln(2)\\ 
S_1\left(\tfrac{1}{8}; \infty\right) &=& - S_{-1}(\infty) + 
S_1\left(-\tfrac{1}{2};\infty\right)~. 
\end{eqnarray}
%----------------------------------------------------------------------------------------------
In various cases S-sums at infinity reduce to MZVs, cf. also \cite{Ablinger:2012qm},
%----------------------------------------------------------------------------------------------
\begin{eqnarray}
S_{1,1,1}\left(\tfrac{1}{2},2,1; \infty \right) &=& \frac{3}{2} \zeta_2 \ln(2) + \frac{7}{4} 
\zeta_3 \\
S_{2,1}\left(\tfrac{1}{2},1; \infty \right)     &=& -\frac{1}{2} \ln(2) \zeta_2 + \zeta_3 \\
S_m\left(\tfrac{1}{2}; \infty \right)           &=& \Li_m\left(\tfrac{1}{2}\right)~.
\end{eqnarray}
%----------------------------------------------------------------------------------------------
Otherwise, new basis elements occur which have both representations in infinite sums and 
iterated integrals. Using the above relations bases for different sets of S-sums at infinity were 
calculated in  \cite{Ablinger:2012qm}.

%%%%%%%%%%%%%%%%%%%%%%%%%%%%%%%%%%%%%%%%%%%%%%%%%%%%%%%%%%%%%%%%%%%%%%%%%%%%%%%%%%%%%%%%%%%%%%% 
\section{Cyclotomic Harmonic Sums and Polylogarithms and their Generalization}
\label{sec:7} 
%%%%%%%%%%%%%%%%%%%%%%%%%%%%%%%%%%%%%%%%%%%%%%%%%%%%%%%%%%%%%%%%%%%%%%%%%%%%%%%%%%%%%%%%%%%%%%%

\vspace{1mm}\noindent
The alphabet of the harmonic polylogarithms (\ref{eq:ALPH1}) contains two differential forms
with denominators, which form the first two cyclotomic polynomials: $(1-x)$ and $(1+x)$. It turns 
out that quantum field theoretic calculations are also related to cyclotomic harmonic 
polylogarithms and sums \cite{EX1}. Cyclotomic polynomials are defined by
%----------------------------------------------------------------------------------------------
\begin{eqnarray}
\Phi_n(x) = \frac{x^n-1}{\ds \prod_{d|n, d < n} \Phi_d(x)}~,~~~~~d,n \in \mathbb{N}_+
\end{eqnarray}
%----------------------------------------------------------------------------------------------
and the generating alphabet reads
%----------------------------------------------------------------------------------------------
\begin{eqnarray}  
\label{eq:alphab}   
\mathfrak{A} =
\left\{\frac{dx}{x}\right\}\cup\left\{\left. \frac{x^l dx}{\Phi_k(x)}\right|k\in\mathbb N_+,0\leq
l<\varphi(k)\right\},
\end{eqnarray}
%----------------------------------------------------------------------------------------------
where $\Phi_k(x)$ denotes the $k$th cyclotomic polynomial \cite{LANG}, and $\varphi(k)$ denotes 
Euler's totient function \cite{TOTIENT}. The Poincar\'{e} iterated integrals over the alphabet 
(\ref{eq:alphab}) are called cyclotomic harmonic polylogarithms, cf. \cite{Ablinger:2011te}. Due 
to the regularity of $1/\Phi_n(x)$ for $x \in [0,1]$, except for $\Phi_1(x)$, no more singularities 
appear beyond those known in the case of the usual harmonic polylogarithms (or Nielsen integrals). 
Cyclotomic harmonic polylogarithms obey shuffle relations, cf.~Sect.~\ref{sec:4}.

The cyclotomic harmonic sums \cite{Ablinger:2011te} are related to the cyclotomic harmonic 
polylogarithms via a Mellin transform (\ref{eq:MEL}). The generalized cyclotomic harmonic sums  
are given by
%----------------------------------------------------------------------------------------------
\begin{eqnarray}   
\label{eq:MSU} 
&& S_{\{a_1,b_1,c_1\}, ...,\{a_l,b_l,c_l\}}(s_1, ...,s_l; N)
= \nonumber\\
&& \sum_{k_1 = 1}^{N} \frac{s_1^k}{(a_1 k_1 + b_1)^{c_1}}
S_{\{a_2,b_2,c_2\}; ...;\{a_l,b_l,c_l\}}(s_2, ...,s_l; k_1), 
S_{\emptyset} = 1,
\end{eqnarray}
%----------------------------------------------------------------------------------------------
where $a_i, c_i \in \mathbb{N}_+, b_i \in  \mathbb{N},~~s_i \in \mathbb{R} \backslash \{0\}, a_i > 
b_i$;
the weight of this sum is defined by $c_1+\dots+c_l$ and $\{a_i, b_i, c_i\}$  denote lists, not 
sets. If $s_i = \pm 1$ these are the usual cyclotomic harmonic sums.
The simplest cyclotomic sums are the single sums
%----------------------------------------------------------------------------------------------
\begin{eqnarray}   
\label{eq:MSU1} 
S_{\{a_1,b_1,c_1\}}(\pm 1;N) = \sum_{k=1}^{N} \frac{(\pm 1)^k}{(a_1 k + b_1)^{c_1}}~, 
\end{eqnarray}
%----------------------------------------------------------------------------------------------
i.e. harmonic sums with cyclic gaps in the summation. The cyclotomic harmonic sums obey 
quasi-shuffle relations $(A)$. 

Beyond this the cyclotomic harmonic sums obey structural relations implied by differentiation for 
the upper summation bound $N$, $(D)$, which require to also consider their values at $N \rightarrow 
\infty$. There are, furthermore, multiple argument relations, cf. \cite{Ablinger:2011te}, decomposing 
$S_{a_i,b_i,c_i}(k \cdot N)$, called synchronization $(M)$, and two duplication relations $(H_1,H_2)$. 
Let us consider the cyclotomic harmonic sums implied by the letters
%----------------------------------------------------------------------------------------------
\begin{eqnarray}   
\frac{1}{k^{l_1}},~~
\frac{(-1)^k}{k^{l_2}},~~
\frac{1}{(2k+1)^{l_3}},~~
\frac{(-1)^k}{(2k+1)^{l_4}}~.
\end{eqnarray}
%----------------------------------------------------------------------------------------------
The length of the basis can be calculated by
%----------------------------------------------------------------------------------------------
\begin{eqnarray}   
N_S(w) &=& 4 \cdot 5^{w-1}
\\
N_A(w) &=& \frac{1}{w} \sum_{d|w} \mu\left(\frac{w}{d}\right) 5^d
\\
N_D(w) &=& N_S(w) - N_S(w-1)
\\
N_{A,D}(w) &=& N_A(w) - N_A(w-1)
\\
N_{A,D,M,H_1,H_2}(w) &=& 
  \frac{1}{w} \sum_{d|w} \mu\left(\frac{w}{d}\right) (5^2 - 3 \cdot 2^d)
- \frac{1}{w-1} \sum_{d|w-1} \mu\left(\frac{w-1}{d}\right) (5^2 - 3 \cdot 2^d),
\nonumber\\
\end{eqnarray}
%----------------------------------------------------------------------------------------------
where $N_S(w)$ denotes the number of all sums. One may calculate the asymptotic representation of 
the cyclotomic harmonic sums analytically. Here also the values of cyclotomic sums at 
$N \rightarrow \infty$ occur. The singularities of the cyclotomic harmonic sums with $s_k =  \pm 
1$  are situated at the non-positive integers.

The cyclotomic sums for $N \rightarrow \infty$ are denoted by
$\sigma_{\{a_1,b_1,c_1\}, ...,\{a_l,b_l,c_l\}}(s_1, ...,s_l)$. For $\forall~|s_k| \leq 1$
divergent series do only occur if $a_1 = b_1$ and $c_1 = 1$, where the degree of divergence 
is given by $\sigma_0$ as in the case of the harmonic sums and can be represented algebraically.
They are related to the values of the cyclotomic harmonic polylogarithms at $x=1$.
At $w=1$ the regularized sums may be given in terms of $\psi(k/l)$ and for higher weights
in terms of  $\psi^{(m)}(k/l), m \geq 1$. If $l$ is an integer for which the $l$-polygon 
is constructible one obtains representations in terms of algebraic numbers and logarithms of 
algebraic numbers, as well as $\pi$ \cite{Ablinger:2011te}. In this way, $\zeta_2$ being a 
basis element in case of the MZVs, looses its role. At depth $w=2$ Catalan's constant 
\cite{CATALAN} with
%----------------------------------------------------------------------------------------------
\begin{eqnarray}   
\sigma_{2,1,-2} = -1  + {\bf C}
\end{eqnarray}
%----------------------------------------------------------------------------------------------
contributes. At higher depth new numbers emerge, which partly can be given integral 
representations involving polylogarithms and roots of the integration variable $x$.
The cyclotomic sums at infinity, as real representations, are closely related to the infinite 
generalized harmonic sums at weights $s_k$ which are roots of unity, cf. also \cite{Broadhurst:1998rz}.
In \cite{Ablinger:2011te} basis representations were worked out for $w = 1,2$ for the $l$th roots, 
$l \in [1,20]$, cf. also \cite{RAC}. Counting relations for bases of the cyclotomic sums at infinity 
have also been derived in Ref.~\cite{Ablinger:2011te}.

%%%%%%%%%%%%%%%%%%%%%%%%%%%%%%%%%%%%%%%%%%%%%%%%%%%%%%%%%%%%%%%%%%%%%%%%%%%%%%%%%%%%%%%%%%%%%%% 
\section{Nested Binomial and Inverse-binomial Harmonic Sums and Associated Polylogarithms}
\label{sec:8} 
%%%%%%%%%%%%%%%%%%%%%%%%%%%%%%%%%%%%%%%%%%%%%%%%%%%%%%%%%%%%%%%%%%%%%%%%%%%%%%%%%%%%%%%%%%%%%%%

\vspace{1mm}\noindent
In massive calculations further extensions to the nested sums and iterated integrals being 
discussed in the previous sections occur. Here summation terms  of the kind
$S_{\vec{a},\vec{b},\vec{c}}(\vec{x};k)$, where $\vec{d} = \{d_1,...,d_m\}$, or their linear combinations are 
modulated by
%----------------------------------------------------------------------------------------------
\begin{eqnarray}   
S_{\vec{a},\vec{b},\vec{c}}(\vec{x};k) &\rightarrow& 
\binom{2k}{k} S_{\vec{a},\vec{b},\vec{c}}(\vec{x};k)\nonumber\\
S_{\vec{a},\vec{b},\vec{c}}(\vec{x};k) &\rightarrow& 
\frac{1}{\ds \binom{2k}{k}} S_{\vec{a},\vec{b},\vec{c}}(\vec{x};k)~,
\label{eq:BIN}
\end{eqnarray}
%----------------------------------------------------------------------------------------------
building iterated sums \cite{ABRSW13}. Sums of this kind occur in case of V-type 3--loop graphs 
for massive operator matrix elements. Simpler sums are obtained in case of 3--loop graphs with 
two fermionic lines of equal mass. Single sums of this kind have been considered earlier, see e.g.
\cite{BIN1}. One may envisage generalizations of (\ref{eq:BIN}) in choosing for the binomial a 
general hypergeometric term\index{hypergeometric term}, i.e. a function, the ratio of which by all 
shifts of arguments being rational. The association of the corresponding iterated integrals in the foregoing cases 
has been found easily. Here the situation is more difficult and the functions representing these 
iterated sums are found in establishing differential equations \cite{ABRS13}. It is found in the 
cases occurring in Ref.~\cite{ABRSW13} that the corresponding differential equations finally factorize 
and one obtains iterated integrals over alphabets which also contain root--valued letters
%----------------------------------------------------------------------------------------------
\begin{eqnarray}   
\label{eq:BINS2}
\frac{1}{\sqrt{(x+a)(x+b)}},~~~~\frac{1}{\sqrt{(x+a)(x+b)}}\frac{1}{x+c},~~~a,b,c 
\in \mathbb{Q} 
\end{eqnarray}
%----------------------------------------------------------------------------------------------
beyond those occurring in case of the generalized (cyclotomic) polylogarithms. A few examples
of this type have been considered in \cite{RVHPL1}. The relative transcendence of the 
nested sums and iterated integrals has been proven. The V-type 3--loop graphs require
alphabets of about 30 root-valued letters. The corresponding nested sums do partly 
diverge $\propto a^N, a \in \mathbb{N}, a \geq 2$. A typical example for a nested binomial 
sums is given by~:
%----------------------------------------------------------------------------------------------
\begin{eqnarray}   
&& \sum_{i=1}^N \binom{2i}{i}(-2)^i \sum_{j=1}^i \frac{1}{\ds j \binom{2j}{j}}
S_{1,2}\left(\tfrac{1}{2},-1;j\right) \\
&&  = \int_0^1 dx \frac{x^N-1}{x-1} \sqrt{\frac{x}{8+x}}\left[{\rm H}^*_{w_{17},-1,0}(x) - 2 
{\rm H}^*_{w_{18},-1,0}(x)\right] 
\nonumber\\ &&
+ \frac{\zeta_2}{2} \int_0^1 dx \frac{(-x)^N-1}{x+1} \sqrt{\frac{x}{8+x}}\left[{\rm 
H}^*_{12}(x)
-2 {\rm H}^*_{13}(x)\right]
%\nonumber\\ &&
+ c_3 \int_0^1 dx \frac{(-8x)^N-1}{x+\frac{1}{8}} \sqrt{\frac{x}{1-x}}~,
\nonumber \label{eq:BINS1}
\end{eqnarray}
%----------------------------------------------------------------------------------------------
with $c_3 = \sum_{j=0}^\infty S_{1,2}\left(\tfrac{1}{2},-1;j\right)(j!)^2/j/(2j)!/\pi$ 
one
of the specific constants emerging in case of these sums. Here the iterated integrals 
$\HA^*$ extend to $x=1$ as firm bound, contrary to the cases discussed before where $x=0$
is chosen. Here the new letters $w_k$ are
%----------------------------------------------------------------------------------------------
\begin{eqnarray}   
w_{12} = \frac{1}{\sqrt{x(8-x)}},~~ 
&& w_{13} = \frac{1}{(2-x)\sqrt{x(8-x)}},\nonumber\\
w_{17} = \frac{1}{\sqrt{x(8+x)}},~~
&&w_{18} = \frac{1}{(2+x)\sqrt{x(8+x)}}~.
\end{eqnarray}
%----------------------------------------------------------------------------------------------
The representations over the letters (\ref{eq:BINS2}) are needed to eliminate the power growth $\propto a^N$ of 
these sums and can be used to derive the asymptotic representation at large values of $N$.
While the terms $\propto 8^N$ and $\propto 4^N$ cancel, it may occur that individual scalar diagrams 
exhibit contributions $\propto 2^N$, cf.~\cite{ABRSW13}. This behaviour is expected to cancel in the complete 
physics result.
%%%%%%%%%%%%%%%%%%%%%%%%%%%%%%%%%%%%%%%%%%%%%%%%%%%%%%%%%%%%%%%%%%%%%%%%%%%%%%%%%%%%%%%%%%%%%%%
\section{Analytic Continuation of Harmonic Sums}
\label{sec:9}
%%%%%%%%%%%%%%%%%%%%%%%%%%%%%%%%%%%%%%%%%%%%%%%%%%%%%%%%%%%%%%%%%%%%%%%%%%%%%%%%%%%%%%%%%%%%%%%

\vspace{1mm}\noindent
The loop-corrections to various physical quantities take a particular simple form in Mellin-space
being expressed in terms of harmonic sums and their generalizations. 
Moreover, in this representation 
the renormalization group equations can be solved analytically, cf.~\cite{CASY,Blumlein:2012bf}. 
For a wide variety
of non-perturbative parton distributions Mellin-space representations can be given as well, 
see e.g.~\cite{Blumlein:2011zu}. 

\vspace*{2.7cm}
\hspace*{3.8cm}
%\begin{figure}%\centering
\setlength{\unitlength}{1mm}
\begin{picture}(-30,0)(-10,10)
\put(0,-10){\vector(0,0){40}}
\put(-20,10){\vector(1,0){50}}
\put(0,10){\circle*{1}}
\put(-21,5){\sf \small -4}
\put(-16,5){\sf \small -3}
\put(-11,5){\sf \small -2}
\put(-6,5){\sf \small -1}
\put( 4,5){\sf \small 1}
\put(9,5){\sf \small 2}
\put(14,5){\sf \small 3}
\put(19,5){\sf \small 4}
\put(20,9.5){\line(0,1){1}}
\put(15,9.5){\line(0,1){1}}
\put(10,9.5){\line(0,1){1}}
\put(5,9.5){\line(0,1){1}}
\put(24,5){\sf \small Re(N)}
\put(2,30){\sf \small Im(N)}
\put(-0.5,25){\line(1,0){1}}
\put(-0.5,20){\line(1,0){1}}
\put(-0.5,15){\line(1,0){1}}
\put(-0.5,5){\line(1,0){1}}
\put(-0.5,0){\line(1,0){1}}
\put(-0.5,-5){\line(1,0){1}}
\put(-0.5,-10){\line(1,0){1}}
\put(-5,10){\circle*{1}}
\put(-10,10){\circle*{1}}
\put(-15,10){\circle*{1}}
\put(-20,10){\circle*{1}}
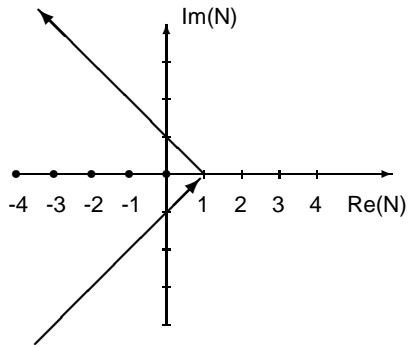
\put(-20,-17){\small {\bf Fig.~2} Path of the contour integral (\ref{eq:CONT}).}
\thicklines
\put(-17.5,-12.5){\vector(1,1){22}}
\put(5,10){\vector(-1,1){22}}
\end{picture}
%\end{figure}

\vspace*{2.3cm}
\noindent
Thus one obtains complete representations for observables
in $N$-space. In case of the perturbative part, the singularities are situated at the integers $N 
\leq N_0$, with usually $N_0 = 1$, see e.g. \cite{Blumlein:1997em}. The harmonic sums possess a 
unique polynomial representation
in terms of the sum $S_1(N)$ and harmonic sums which can be represented as Mellin transforms
having a representation by factorial series \cite{NIELS,LANDAU}.
They are transformed to $x$-space by a single precise numerical contour integral
around the singularities of the problem to compare with the data measured in experiment.
The analytic continuation of the perturbative evolution kernels and Wilson coefficients
from even or odd integers to complex values of $N$ is unique \cite{Carlson:14}\footnote{For a 
detailed proof also in case of generalized harmonic sums see \cite{Ablinger:2013cf}.}. To perform 
this integral a representation of the harmonic sums for $N \in \mathbb{C}$ is required.
Accurate numeric representations up to $w = 5$ have been given in \cite{ANCONT}, see also
\cite{Kotikov:2005gr}. Arbitrary precise representations can be obtained using the analytic
expressions for the asymptotic representation \cite{Blumlein:2009ta,Blumlein:2009fz}
together with the recursion relations given in Sect.~\ref{sec:3}. These are given up to
$w = 8$ in \cite{ABS13}. A path to perform the inverse Mellin transform 
%---------------------------------------------------------------------------------------------
\begin{eqnarray}
\hspace*{-1cm}
f(x) = \frac{1}{\pi} \int_0^{\infty} dz~{\sf Im}\left[
e^{i\phi}~x^{-C} \Mvec[f](N=C)\right],~~~C=c+z e^{i\phi}
\label{eq:CONT}
\end{eqnarray}
%---------------------------------------------------------------------------------------------
is shown in Figure~2. The asymptotic representations can also be obtained in analytic form for the
S-sums \cite{Ablinger:2013cf}, cyclotomic (S)-sums \cite{Ablinger:2011te}, as well as for the 
nested binomial cyclotomic S-sums \cite{ABRSW13,ABRS13}.

In case the expressions in $N$-space result from Mellin transforms of functions $f(x) \sim 
x^\alpha, \alpha \in ]0,1[$ the singularities are shifted by $\alpha$. This is usually the case
for the non-perturbative parton distribution functions, but also in case of some root--valued
harmonic polylogarithms considered in Sect.~\ref{sec:8}.
 
The inverse Mellin-transform cannot be performed in the above way for integrands which do not 
vanish sufficiently fast enough as $|N| \rightarrow \infty$. Contributions of this kind are those leading 
to distribution--valued terms in $x$-space as to $\delta(1-x), [\ln^k(1-x)/(1-x)]_+$ the Dirac 
$\delta$-distribution and the $+$-distribution defined by
%-------------------------------------------------------------------------------------------------
\begin{eqnarray}
\int_0^1~dx~\left[f(x)\right]_+ g(x) = \int_0^1~dx~\left[g(x)-g(1)\right] f(x)~.
\end{eqnarray}
%-------------------------------------------------------------------------------------------------
Also for terms which grow like $a^N, a \in \mathbb{R}, a > 1$ in $N$-space, the Mellin transform
cannot be performed numerically in general. They are not supposed to emerge in physical observables.
The physical quantities in hadronic scattering contain the parton distribution functions, which, however, 
damp according contributions occurring in the evolution kernels sufficiently. On the other hand, 
the inverse Mellin transform can always be performed analytically changing form nested sum-representations 
in $N$-space to iterated integral representations in $x$-space as has been outlined before.
%-------------------------------------------------------------------------------------------------
\setcounter{figure}{2}

\def\circlea{(-8.3cm,6cm) circle (1.8cm and 1cm)}
\def\circleb{(-6.5cm,6cm) circle (1.8cm and 1cm)}
\def\circlec{(-1.8cm,6cm) circle (1.8cm and 1cm)}
\def\circled{(-0.0cm,6cm) circle (1.8cm and 1cm)}
\def\circlee{(-8.3cm,2cm) circle (1.8cm and 1cm)}
\def\circlef{(-6.5cm,2cm) circle (1.8cm and 1cm)}
\def\circleg{(-1.8cm,2cm) circle (1.8cm and 1cm)}
\def\circleh{(-0.0cm,2cm) circle (1.8cm and 1cm)}

\tikzset{filled/.style={fill=circle area, draw=circle edge, thick}, outline/.style={draw=circle edge, thick}}

\colorlet{circle edge}{black!100}
\colorlet{circle area}{black!20}
\setlength{\parskip}{5mm}
\begin{figure}[t!]
\centering
\begin{tikzpicture}
\centering
     \begin{scope}
         \clip \circlea;
         \fill[filled] \circleb;
     \end{scope}
     \draw[outline] \circlea;
     \draw[outline] \circleb;
     \draw (-7.40cm,6.3cm) node[font=\small] {H-Sums};
     \draw (-7.35cm,5.7cm) node[font=\tiny] {$\S{-1,2}n$};
     \draw (-9.05cm,6.3cm) node[font=\small] {S-Sums};
     \draw (-9.05cm,5.7cm) node[font=\tiny] {$\S{1,2}{\frac{1}{2},1;n}$};
     \draw (-5.70cm,6.3cm) node[font=\small] {C-Sums};
     \draw (-5.65cm,5.7cm) node[font=\tiny] {$\S{(2,1,-1)}n$};
     \begin{scope}
         \clip \circlec;
         \fill[filled] \circled;
     \end{scope}
     \draw[outline] \circlec;
     \draw[outline] \circled;
     \draw (-0.90cm,6.3cm) node[font=\small] {H-Logs};
     \draw (-0.85cm,5.7cm) node[font=\tiny] {$\H{-1,1}x$};
     \draw (-2.55cm,6.3cm) node[font=\small] {C-Logs};
     \draw (-2.60cm,5.7cm) node[font=\tiny] {$\textnormal{H}_{(4,1),(0,0)}\hspace{-0.05cm}(x)$};
     \draw (+0.80cm,6.3cm) node[font=\small] {G-Logs};
     \draw (+0.85cm,5.7cm) node[font=\tiny] {$\H{2,3}x$};

     \draw[thick,->] (-9cm,6.5cm) .. controls +(80:1cm) and +(100:1cm) .. (1cm,6.5cm) node[midway,sloped,above,font=\small] {integral representation (inv. Mellin transform)};
     \draw[thick,->] (-7.25cm,6.5cm) .. controls +(60:1cm) and +(120:1cm) .. (-0.75cm,6.5cm);
     \draw[thick,->] (-5.50cm,6.5cm) .. controls +(40:1cm) and +(140:1cm) .. (-2.5cm,6.5cm);

     \draw[thick,<-] (-9cm,5.5cm) .. controls +(280:1cm) and +(260:1cm) .. (1cm,5.5cm) node[midway,sloped,below,font=\small] {Mellin transform};
     \draw[thick,<-] (-7.25cm,5.5cm) .. controls +(300:1cm) and +(240:1cm) .. (-0.75cm,5.5cm);
     \draw[thick,<-] (-5.5cm,5.5cm) .. controls +(320:1cm) and +(220:1cm) .. (-2.5cm,5.5cm);

     \begin{scope}
         \clip \circlee;
         \fill[filled] \circlef;
     \end{scope}
     \draw[outline] \circlee;
     \draw[outline] \circlef;
     \draw (-7.40cm,2cm) node[font=\tiny] {$\S{-1,2}\infty$};
     \draw (-9.15cm,2cm) node[font=\tiny] {$\S{1,2}{\frac{1}{2},1;\infty}$};
     \draw (-5.60cm,2cm) node[font=\tiny] {$\S{(2,1,-1)}\infty$};
     \draw[thick,->] (-7.4cm,4.7cm) -- (-7.4cm,3.3cm) node[midway,sloped,above,font=\small] {$n\rightarrow\infty$};

     \begin{scope}
         \clip \circleg;
         \fill[filled] \circleh;
     \end{scope}
     \draw[outline] \circleg;
     \draw[outline] \circleh;
     \draw (-0.85cm,2cm) node[font=\tiny] {$\H{-1,1}1$};
     \draw (-2.65cm,2cm) node[font=\tiny] {$\textnormal{H}_{(4,1),(0,0)}\hspace{-0.05cm}(1)$};
     \draw (0.85cm,2cm) node[font=\tiny] {$\H{2,3}c$};
     \draw[thick,->] (-0.9cm,5.3cm) -- (-0.9cm,2.7cm) node[midway,sloped,above,font=\small] {$x\rightarrow 1$};
     \draw[thick,->] (-2.0cm,5.3cm) -- (-2.0cm,2.7cm) node[midway,sloped,above,font=\small] {$x\rightarrow 1$};
     \draw[thick,->] (0.2cm,5.3cm) -- (0.2cm,2.7cm) node[midway,sloped,above,font=\small] {$x\rightarrow c\in \mathbb R$};

     \draw[thick,<-] (-9cm,1.5cm) .. controls +(280:1cm) and +(260:1cm) .. (1cm,1.5cm) node[midway,sloped,below,font=\small] {power series expansion};
     \draw[thick,<-] (-7.25cm,1.5cm) .. controls +(300:1cm) and +(240:1cm) .. (-0.75cm,1.5cm);
     \draw[thick,<-] (-5.5cm,1.5cm) .. controls +(320:1cm) and +(220:1cm) .. (-2.5cm,1.5cm);

     \draw[thick,->] (-9cm,2.5cm) .. controls +(80:1cm) and +(100:1cm) .. (1cm,2.5cm);
     \draw[thick,->] (-7.25cm,2.5cm) .. controls +(60:1cm) and +(120:1cm) .. (-0.75cm,2.5cm);
     \draw[thick,->] (-5.50cm,2.5cm) .. controls +(40:1cm) and +(140:1cm) .. (-2.5cm,2.5cm);

\end{tikzpicture}
\caption{\label{Ifig1}Connection between harmonic sums (H-Sums), S-sums (S-Sums) and cyclotomic harmonic sums (C-Sums), their values at infinity and harmonic polylogarithms (H-Logs), generalized harmonic polylogarithms (G-Logs) and 
cyclotomic harmonic polylogarithms (C-Logs) and their values at special constants.}
\end{figure}
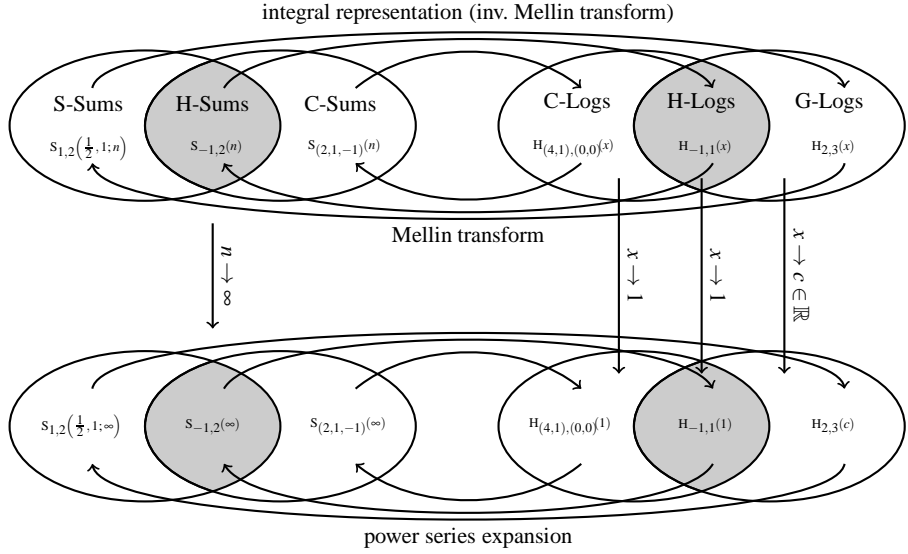
%-------------------------------------------------------------------------------------------------
%%%%%%%%%%%%%%%%%%%%%%%%%%%%%%%%%%%%%%%%%%%%%%%%%%%%%%%%%%%%%%%%%%%%%%%%%%%%%%%%%%%%%%%%%%%%%%%
\section{Conclusions}
\label{sec:10}
\setcounter{figure}{2}
%%%%%%%%%%%%%%%%%%%%%%%%%%%%%%%%%%%%%%%%%%%%%%%%%%%%%%%%%%%%%%%%%%%%%%%%%%%%%%%%%%%%%%%%%%%%%%%

\vspace{1mm}\noindent
Feynman integrals in Quantum Field Theories generate a hierarchic series of special functions,
which allow their unique representation. They emerge in terms of special nested sums, iterated
integrals and numbers. Their variety gradually extends enlarging the number of loops and legs, as
well as the associated mass scales. The systematic exploration of these structures has been 
started about 15 years ago and several levels of complexity have been unraveled since. The 
relations of the various associated sums and integrals are schematically illustrated in Fig.~3 and 
are widely explored. Many relations are implied by the shuffle resp. stuffle algebras, others 
are structural relations. The number of relations grows with the number of admissible 
operations.

A large amount of transformations and relations between the different quantities being discussed
in this article are encoded in the package {\tt HarmonicSums} 
\cite{Ablinger:2013cf,JA1} for public use.
For newly emerging structures the algebraic relations are easily generalized but they will 
usually apply structural relations of a new kind. With the present programme revealing their 
strict (atomic) structure, they are fully explored analytically and Feynman's original approach
to completely organize the calculation of observables in Quantum Field Theory is currently extended
to massive calculations at the 3-loop level in Quantum Electrodynamics and Quantum Chromodynamics
at the perturbative side. For these quantities efficient numerical representations 
have to be derived. Working in Mellin space the treatment may even remain {\it completely analytic},
in a very elegant way, up to a single final numerical contour integral around the singularities 
of the problem, cf.~Sect.~\ref{sec:9}.
 
Despite of the achievements being obtained many more physical classes still await their systematic
exploration in the future. It is clear, however, that the various concrete structures are realized 
as combinations of words over certain alphabets, which may be called the {\it genetic code of the 
microcosm}~\cite{BRS}. 
%%%%%%%%%%%%%%%%%%%%%%%%%%%%%%%%%%%%%%%%%%%%%%%%%%%%%%%%%%%%%%%%%%%%%%%%%%%%%%%%%%%%%%%%%%%%%%%

\vspace*{2mm}\noindent
{\bf Acknowledgment.}~We would like to thank 
D.~Broadhurst,
F.~Brown,
A.~De Freitas,
E.W.N.~Glover,
A.~Hasselhuhn,
D.~Kreimer,
C.~Raab,
C.~Schneider,
S.~Weinzierl,
F.~Wi\ss{}brock,
and J.~Vermaseren for discussions.
This work has been supported in part by DFG
Sonderforschungsbereich Transregio 9, Computergest\"utzte Theoretische
Teilchenphysik, by the Austrian Science Fund (FWF) grants P20347-N18, P22748-N18, SFB F50 (F5009-N15)
and by the EU Network {LHCPHENOnet}~PITN-GA-2010-264564.

%%%%%%%%%%%%%%%%%%%%%%%%%%%%%%%%%%%%%%%%%%%%%%%%%%%%%%%%%%%%%%%%%%%%%%%%%%%%%%%%%%%%%%%%%%%%%%%

%%%%%%%%%%%%%%%%%%%%%%%%%%%%%%%%%%%%%%%%%%%%%%%%%%%%%%%%%%%%%%%%%%%%%%%%%%%%%%%%%%%%%%%%%%%%%%%
\end{document}